\documentclass[12pt,preprint]{aastex}

\newcommand{\bdv}[1]{\mbox{\boldmath$#1$}}

\def\au{{\rm au}}

\def\masyr{{\rm mas}\,{\rm yr}^{-1}}
\def\kpc{{\rm kpc}}
\def\mas{{\rm mas}}

\def\muas{\mu{\rm as}}

\def\max{{\rm max}}
\def\min{{\rm min}}
\def\rel{{\rm rel}}

\def\eff{{\rm eff}}

\def\e{{\rm E}}
\def\bpi{{\bdv\pi}}
\def\bmu{{\bdv\mu}}

\def\btheta{{\bdv\theta}}

\begin{document}
\title{Industrial-Scale Mass Measurements of Isolated Black Holes}

\author{\textsc{
Andrew Gould$^{1,2}$}}

\affil{$^{1}$Max-Planck-Institute for Astronomy, K\"{o}nigstuhl 17,
69117 Heidelberg, Germany}

\affil{$^{2}$Department of Astronomy, Ohio State University, 140 W.
18th Ave., Columbus, OH 43210, USA}

\begin{abstract}

  I show that industrial-scale mass measurement of isolated black
  holes (BHs) can be achieved by combining a high-cadence, wide-field
  microlensing survey such as KMTNet, observations from a parallax
  satellite in solar orbit, and VLTI GRAVITY+ interferometry.  I show
  that these can yield precision measurements of microlens parallaxes
  down to $\pi_\e\sim 0.01$ and Einstein radii down to $\theta_\e\sim
  1\,\mas$.  These limits correspond to BH masses $M\sim 12\,M_\odot$,
  deep in the Galactic bulge, with lens-source separations of
  $D_{LS}\sim 0.6\,$kpc, and they include all BHs in the Galactic
  disk.  I carry out detailed analyses of simulations that explore
  many aspects of the measurement process, including the decisions on
  whether to carry out VLTI measurements for each long-event
  candidate.  I show that the combination of ground-based and
  space-based light curves of BH events will automatically exclude the
  spurious ``large parallax'' solutions that arise from the standard
  \citep{refsdal66} analysis, except for the high-magnification
  events, for which other methods can be applied.  The remaining two-fold
  degeneracy can always be broken by conducting a second VLTI
  measurement, and I show how to identify the relatively rare cases
  that this is required.

\end{abstract}


\section{{Introduction}
\label{sec:intro}}

Microlensing is the only known method to measure the masses of isolated
dark objects, in particular isolated black holes (BHs).  Such measurements
require the determination of two parameters that are not returned by
standard \citep{pac86} fits to the microlensing light curve, i.e., the
Einstein radius, $\theta_\e$, and the microlens parallax $\bpi_\e$,
\begin{equation}
  \theta_\e \equiv \sqrt{\kappa M \pi_\rel},\qquad
  \bpi_\e \equiv {\pi_\rel\over\theta_\e}{\bmu_\rel\over\mu_\rel}; \qquad
  \kappa\equiv {4 G\over c^2\au}= 8.14\,{\mas\over M_\odot}.
  \label{eqn:pie-thetae}
\end{equation}
Here, $M$ is the lens mass while $\pi_\rel$ and $\bmu_\rel$ are the
lens-source relative parallax and proper-motion.  Thus \citep{gould92,gould00},
\begin{equation}
M = {\theta_\e\over\kappa\pi_\e},
  \label{eqn:m-eval}
\end{equation}
where $\pi_\e=|\bpi_\e|$.

However, although
this principle has been known for three decades, to date,
there has been only one isolated BH mass measurement
\citep{ob110462a,ob110462b,ob110462c}, despite the fact that of order 1\% of
the $\sim 3\times 10^4$ known microlensing events are likely due to BHs.
The main problem is that for dark objects, measurements of $\theta_\e$ and
$\bpi_\e$ are each rare, so that their overlap is extremely rare.
See \citet{kb222397} for a systematic discussion.

For such isolated dark objects, there are two established methods to
measure $\bpi_\e$ and three established methods to measure $\theta_\e$.
In brief, $\bpi_\e$ can be measured either by observing the event
from a single site on an accelerated platform \citep{gould92} or
by simultaneous observations from two well-separated observatories
\citep{refsdal66}.  Of course, almost all microlensing events are
observed from Earth, which is an accelerated platform, whose parallactic
motion induces annual bumps in the light curve, with an amplitude that is
proportional to $\pi_\e$ and a phase that reflects the direction of $\bmu_\rel$.
However, as most events are short compared to a year, it is generally
difficult to extract $\bpi_\e$.  See Figure~1 of \citet{gouldhorne}.
While BHs tend to have longer timescales $t_\e=\theta_\e/\mu_\rel$
(because $\theta_\e\propto M^{1/2}$), typical BH events are still short
compared to a year.  Moreover, typical BHs have small $\pi_\e\propto M^{-1/2}$.
For example, it is likely that more than half of BH events are due to bulge
BHs, which have $\pi_\rel\sim {\cal O}(15\,\muas)$, so that an $M=10\,M_\odot$
BH would have $\pi_\e\sim 0.014$.  Hence, a $5\,\sigma$ (20\%) measurement
would require $\sigma(\pi_\e)= 0.003$, which (for typical proper motions,
$\mu_\rel \sim 6\,\masyr$, i.e.,
$t_\e=\theta_\e/\mu_\rel = \kappa M \pi_\e/\mu_\rel \sim 70\,$days), are
extremely rare.  Moreover, systematic errors due to unmodeled instrumental
effects in the light curve are believed to be significantly larger than this
threshold.  In particular, the one measured BH, OGLE-2011-BLG-0462, lies
far in the opposite extreme of parameters space, a very rare event with
$\pi_\rel \sim 0.5\,\mas$.

Hence, if it can be made into a practical option, the other approach, i.e.,
observations by a second observatory would be much preferred.  In this case, the
parallax is given approximately by
\begin{equation}
  \bpi_\e = {\au\over D_\perp}\biggl({\Delta t_0\over t_\e},\Delta u_0\biggr)
\label{eqn:sat-parallax}
\end{equation}
where ${\bf D}_\perp$ is the vector separation of the two observatories,
$\Delta t_0$ is the difference of times of peak ($t_0$) between
these observatories, and $\Delta u_0$ is their difference of
impact parameters ($u_0$) normalized to $\theta_\e$.  The two components
are in the directions parallel and perpendicular to ${\bf D}_\perp$.
As already recognized by \citet{refsdal66}, Equation~(\ref{eqn:sat-parallax})
has a 4-fold ambiguity because $u_0$ is a signed quantity, but only
its amplitude is typically measured from \citet{pac86} fits to the light curve.
See Figure~1 from \citet{gould94b}.
This can be a serious issue, but I defer discussion of it to the exposition of
the method that I propose.  I should also mention (but also defer discussion of)
another issue that impacts this measurement via the second component
of Equation~(\ref{eqn:sat-parallax}): it is subject to much larger
statistical errors than the first component \citep{gould95}, as well as
much larger systematic errors.  In addition, for the most practical
implementation of this idea, i.e., a satellite in solar orbit with
$D_\perp\sim 1\,\au$, and for bulge lenses, the typical values of the
two components will be small, $|\Delta t_0/t_\e|\la 0.01$ and
$|\Delta u_0|\la 0.01$, so that control of statistical and systematic errors
can be a major issue.

By far, the most common method to measure $\theta_\e$ for isolated objects,
including dark ones, has been from so-called ``finite-source'' (FS) effects
for the special case that the lens transits the source
\citep{gould94a,witt94,nemiroff94}.  This includes the first
isolated dark-object mass measurement,
OGLE-2007-BLG-0224 \citep{ob07224}, which was a brown dwarf
(BD, \citealt{ob07224b}).  Indeed, of the 30 $\theta_\e$ measurements of
isolated objects with giant-star sources from the systematic study by
\citet{gould22}, of order 1/3 were dark, including four free-floating planet
(FFP) candidates and of order six or more BDs.  However, for the same
reason that this is a powerful method for $\theta_\e$ measurements of low-mass
dark objects, it is almost useless for BHs.  That is, the probability of
such transits is $p\simeq \min(1,\rho)$, where $\rho\equiv \theta_*/\theta_\e$
and $\theta_*$ is the angular radius of the source.  Hence, of the several
hundred isolated-BH events that have likely occurred, we expect less than one to
have shown FS effects.

A second method to measure $\theta_\e$ is astrometric microlensing
\citep{walker95,hnp95,my95}, and indeed this was the method applied for
the only isolated BH mass measurement, i.e., of OGLE-2011-BLG-0462.
The method relies on the fact that the centroid of light from the
two microlensed images is displaced from the source by
\begin{equation}
  \Delta \btheta_{\rm cent} = {{\bf u}\over u^2 + 2}\theta_\e,
  \label{eqn:astrometic}
\end{equation}
where ${\bf u}$ is the vector separation of the source relative to the
lens, normalized to $\theta_\e$.  In particular, the amplitude $u =
|{\bf u}|$ is known from the instantaneous magnification
$A=(u^2+2)/(u\sqrt{u^2+4})$ \citep{einstein36}, i.e., $u^2=
2[(1-1/A^2)^{-1/2}-1]$.  After the source proper motion $\bmu_S$ is
measured from late-time astrometric observations and is subtracted
out, the light centroid traces an ellipse whose semi-major axis is
$\theta_a = \theta_\e/\sqrt{8}$ (provided that $u_0\leq \sqrt{2}$).
This method is challenging, in part because the observations must be
carried out for many $t_\e$ in order to measure $\bmu_S$, and in part
because the amplitude of the effect is small given present astrometric
technology at the relatively faint magnitudes of typical microlensing
events.  That is, as mentioned above, typical bulge BHs have
$\theta_\e=\kappa M\pi_\e\sim 1.1\,\mas$, implying $\theta_a\sim
300\,\muas$.  Hence, a 20\% measurement would require an error of just
$\sigma(\theta_a)\sim 60\,\muas$.  In fact, the $\sim 10\%$ measurement
of $\theta_\e$ for the case of OGLE-2011-BLG-0462 was only possible
because $\theta_\e\sim 5.6\,\mas$ was exceptionally large, which again
was due to fact that the BH was unusually close to the Sun.  Thus,
with present technology, this technique cannot be reliably applied to
bulge BHs and presents considerable challenges for typical disk BHs,
which have $\theta_\e\la 2\,\mas$.

The third method is to resolve the two microlensed images using interferometry
\citep{delplancke01,kojima1,cassan21}.  Their separation is given by
\begin{equation}
  \Delta \btheta_{-,+}=2\sqrt{1 + {u^2\over 4}}\theta_\e{\bf\hat u},
  \label{eqn:interometric}
\end{equation}
Note that the square-root term is very nearly unity, i.e., $\sim 1 + u^2/8$,
and in any case is known very precisely.  Hence, if this method can be
applied at all, it gives extremely precise determinations of $\theta_\e$.
There are three restrictions on such applications, one of principle and the
other two practical.  The restriction of principle is that for any given
baseline configuration of the interferometer, there is a threshold of
detection.  For the only interferometer capable of reaching typical
microlensing events, i.e., the Very Large Telescope Interferometer (VLTI),
this threshold is separations of about 2 mas, i.e., $\theta_\e\ga 1\,\mas$.
Fortunately, this threshold allows measurements for essentially all
disk BHs and for half or more of bulge BHs.

The first practical consideration is that the sensitivity of the VLTI
depends critically on engineering issues.  Hence, for the original VLTI
GRAVITY instrument \citep{gravity}, only targets that are much brighter than
typical microlensing events were accessible.  This is the reason that
the first successful resolution of microlensed images was for an event
that was several hundred times brighter than typical \citep{kojima1}.
However, with the advent of ``GRAVITY Wide'' \citep{gravitywide}, of
order 10\% of microlensing events are already accessible.  Moreover, a future
upgrade to ``GRAVITY+'' \citep{gravity+} is already in progress, after which the
bulk of microlensing events will be accessible.

The second practical consideration is that observing time must be
allocated to measuring BH masses.  Of course, this requirement is
quite generic to any observational project that requires more
expensive instruments than an amateur-class telescope.  However, in
the present case, the issue is potentially massive use of an extremely
sought-after instrument.  In particular, at present, events are
selected for potential VLTI observations based primarily on their
Einstein timescales $t_\e$ because BHs typically have longer $t_\e$.
However, it is straightforward to show that even the longest $t_\e$
events are dominated by lenses with slow $\mu_\rel$ rather than large
$M$ \citep{mao96,kb162052}.  Hence, a substantial majority of BH
candidates that garner VLTI observations turn out to be ordinary
stars.  For pilot programs that are exploring this technology, this is
just the ``cost of doing business''.  However, it would be a major
impediment to an industrial-scale program of BH mass measurements.

\section{{Industrial-Scale Mass Measurements}
\label{sec:industry}}

Based on the overview presented in Section~\ref{sec:intro}, it is clear
that the only feasible path toward industrial-scale BH mass measurements
based on current technologies
is to combine parallax-satellite measurements of $\bpi_\e$ with
VLTI interferometric measurements of $\theta_\e$.  Nevertheless, as is
also clear from the description of these two techniques, several
challenges remain to be addressed.

Before continuing, I should mention that \citet{gouldyee14} proposed a method
based on a future, so far untested, technology.   In Section~\ref{sec:discuss},
I briefly describe this approach and compare it to the one presented here.

\subsection{{Resolution of the Four-Fold Degeneracy}
\label{sec:fourfold}}

The first obvious challenge is the four-fold degeneracy in $\bpi_\e$
from satellite-based measurements.  It is customary to designate these as
$(+,+)$, $(+,-)$, $(-,+)$, and $(-,-)$, where the first entry is the sign
of $u_{0,\rm ground}$ and the second is the sign of $u_{0,\rm sat}$.
Hence, there is one pair [$(+,+)$ \& $(-,-)$] with the same smaller parallax
$\pi_{\e,\rm small}$ (but different directions) and another
pair [$(+,-)$ \& $(-,+)$] with the same larger parallax
$\pi_{\e,\rm large}$.  As I illustrate
below, the different $\pi_\e$ solutions can often be distinguished
from the ground-based light curve, but the different directions
cannot.  Thus, the most critical questions (i.e., is this a BH or not?; and
if it is, what is its mass?) can often be resolved despite the nominal
four-fold degeneracy.  However, the direction
is also important because it is needed to determine the velocity of the BH
relative to its local standard of rest (LSR), which is critical to
mapping out the kick velocities received by the BHs.

Consider an example, for which $\Delta t_0/t_\e=0.01$,
$u_{0,\rm ground}=0.29$, $u_{0,\rm sat}=0.31$, and $D_\perp=1\,\au$.  Then,
the alternate solutions would have $\pi_{\e,\rm small}\sim0.022$ and
$\pi_{\e,\rm large}=0.6$.  For the first case, the impact of the parallax on
the ground-based light curve would likely be unmeasurably small, while
for the second, the unusually large $\pi_\e$ would likely leave clear traces.
Nevertheless, for higher-magnification (lower $u_0$) events, the problems
of distinguishing the solutions would grow.  Moreover, as mentioned above,
these type of arguments do nothing to resolve the directional ambiguity.

This first challenge is automatically addressed by the VLTI
measurement of $\theta_\e$.  As summarized in Section~\ref{sec:intro},
this measurement yields not only the precise image separation (and so
$\theta_\e$), but also the precise direction of the instantaneous
lens-source separation, $\psi$, which is measured relative to north.
This direction differs from that of
$\bmu_\rel$ (and therefore also of $\bpi_\e$) by \citep{kojima1}
\begin{equation}
  \phi=\cot^{-1} {\delta t\over t_\eff}; \qquad t_\eff \equiv |u_0| t_\e;
  \qquad \delta t \equiv t_{\rm obs} - t_0.
  \label{eqn:phidef}
\end{equation}
As I will describe below, both $t_0$ and $t_\eff$ are generally measured
very precisely because they are nearly uncorrelated with other parameters
in the ground-based light-curve fit.  From a single interferometric measurement,
one does not know the sign of this offset because one does not know
the sign of $u_0$.  Thus, to this point, there are two possible directions
of $\bpi_\e$, i.e.,
\begin{equation}
  \Phi_\pi^{u_0 \pm} = \pi + \psi \mp \phi.
  \label{eqn:Phidef}
\end{equation}
Note that the appearance of the transcendental ``$\pi$'' in this equation is
due to the fact that standard microlensing conventions specify that $\bmu_\rel$
is the motion of the lens relative to the source, while interferometry is most
conveniently described using angles relative to the major image, which
is in the same direction as the source relative to the lens.  See Figure~4
from \citet{kojima1}.

A secure method to resolve this ambiguity is simply to make a second
interferometric measurement.  Expressed in terms of Figure~4 from
\citet{kojima1}, if $u_0>0$, then the orientation $\phi$ will move clockwise,
whereas if $u_0<0$, it will move counter-clockwise.  

\subsection{{Minimizing the Number of Interferometric Observations}
\label{sec:minimize-inerferometry}}

A more difficult challenge arises from the sheer volume of interferometric
observations that are required for industrial-scale BH mass measurements.
Of course, there must be at least one interferometric observation for each
successful BH mass measurement, but the problem is that if candidates are
selected, as in current practice, primarily based on relatively long $t_\e$,
then there will be of order 10 targets for each BH mass measurement.
Thus, if two interferometric measurements are required for each measurement
(as anticipated in the previous paragraph), then about 20 interferometric
observations would be required for each BH mass measurement.  Finally,
the present selection procedures are strongly biased against BHs that have
been kicked to high velocity because these have higher proper motions and
so short $t_\e=\theta_\e/\mu_\rel$.  Yet, if the timescale threshold were
relaxed, it would lead to even greater contamination of the sample by
ordinary stars.

\subsubsection{{Single-Epoch Interferometric Observations}
\label{sec:single-epoch}}

I begin by showing that when $\bpi_\e$ is derived from parallax-satellite
observations, it is usually unnecessary to make two interferometric
observations to resolve the degeneracies.  At first, it appears that there
are eight possible overlaps between the two possible directions from
the interferometric observation and the four possible directions from the
parallax-satellite observations.  However, in fact, there are only
four possible overlaps.  That is, for the case $u_0>0$, there is just one
possible interferometric direction and two possible parallax-satellite
directions, so two possible overlaps.  Then, there are also two other
overlaps for the case that $u_0<0$.  The second point is that if the errors
of each of the four $\bpi_\e$ solutions are small compared to their values,
then the angular uncertainty will be small.  Because the angular error
in the $\Phi_\pi^{u_0\pm}$ measurements are negligibly small, this means that
the chance that some pair of directions (other than the actual direction)
will coincide within errors is very small.  Furthermore, unless $|u_0|\ll 1$,
the actual direction will be substantially different from that of the other
solution having the same sign of $u_0$ (say $u_0>0$ for definiteness).
Hence, the only possibility of confusion would be that the $u_0<0$
interferometric direction coincided with one of the two parallax-satellite
$u_0<0$ directions.  Finally, as noted above, in most cases that the
lens is actually a BH (and so $\pi_\e$ is small), the $\pi_{\e,\rm large}$
solutions can be ruled out by the absence of strong parallax signatures
in the ground-based light curve.

If the errors in the four $\bpi_\e$ solutions are not all small compared
to their values, then the situation must be considered more closely.
To be concrete, I assume that the error in component parallel to ${\bf D}_\perp$
(i.e., in the $\Delta t_0/t_\e$ direction) is 0.003, while the error in the
perpendicular component (i.e., the $\Delta u_0$ direction) is 0.01.
The reasons for the asymmetric errors will be discussed below.  And I will
assume a parallax near the threshold of VLTI detection of
$\pi_\e = \theta_\e/\kappa M = (1\,\mas) /\kappa (10\,M_\odot) = 0.012$,
or explicitly $\Delta t_0/t_\e=0.009$, $u_{0,\rm ground}=0.296$,
and $u_{0,\rm sat}=0.304$.  Thus, it is still the case that $|\Delta u_0|=0.6$
for the two $\pi_{\e,\rm large}$ cases, so that the $2\,\sigma$ error ellipse
only subtends about $1.2^\circ$, i.e., 1/300 of the unit circle.  Hence,
it is extremely unlikely that if, for example, the $(+,+)$ solution is correct,
the $u_0<0$ interferometric direction vector would intersect the $(-,+)$
solution.  Nevertheless, in this example, the $2\,\sigma$ error ellipses of
each of the $\pi_{\e,\rm small}$ solutions would subtend $\sim 125^\circ$,
implying a 35\% chance that there would be a second (i.e., spurious) solution
for the case of a single interferometric observation.

Thus, there would be some subset of cases that would require a second
observation to resolve the directional degeneracy between the two
$\pi_{\e,\rm small}$ solutions, and a much smaller subset for which
breaking the more critical large/small $\pi_\e$ degeneracy would
require two interferometric observations.  In Section~\ref{sec:examples},
I describe how these observational decisions could be made dynamically.
However, from the present perspective, the main takeaway is that there needs to
be an average of only slightly more than one interferometric
observation per target.

\subsubsection{{Vetting Against Contaminants}
\label{sec:vetting}}

The next challenge is to limit the number of targets without
substantially reducing the number of BHs.  The central idea for doing
so is to make the first (and likely only) interferometric observation
somewhat after $t_{0,\rm ground}$, when $t_{0,\rm sat}$ and $u_{0,\rm
  sat}$ are approximately measured.  This will, as usual, lead to a
four-fold degenerate measurement of $\bpi_\e$, with two possible
values of $\pi_\e$, i.e., $\pi_{\e,\rm small}$ and $\pi_{\e,\rm large}$.
Next, one would conservatively adopt $\pi_\e = \pi_{\e,\rm small}$.
That is, if $\pi_{\e,\rm small}$ is rejected by the selection criterion
derived below then $\pi_{\e,\rm large}$ would also be rejected,
simply because it is larger.

From this estimate of $\pi_\e$ and the measured value of $t_\e$, which
will be quite well determined by this time, one can predict the
lens-source relative proper motion as a function of the (still unknown)
lens mass $M$,
\begin{equation}
  \mu_\rel = {\theta_\e\over t_\e} = {\kappa M\pi_\e\over t_\e}.
  \label{eqn:m-murel}
\end{equation}
Because the great majority of contaminating lenses will have $M\leq M_\odot$,
this implies that for these contaminants,
$\mu_\rel \leq \kappa M_\odot \pi_\e/t_\e$.
Thus, the indicated way to screen against contamination is to set some
proper-motion threshold, $\mu_{\rm thresh}$, above which one is not willing
to accept contamination.  Then, the criterion for making an interferometric
observation will be
\begin{equation}
  \pi_\e  < {\mu_{\rm thresh} t_\e\over \kappa M_\odot}
  = {t_\e\over 8.14\,{\rm yr}}\,{\mu_{\rm thresh} \over 1\,\masyr}.
  \label{eqn:pie-te}
\end{equation}
For example, if $\mu_{\rm thresh} = 1.5\,\masyr$, then an interferometric
observation will be made provided that $\pi_\e < 0.03(t_\e/60\,{\rm day})$.

In choosing $\mu_{\rm thresh}$, one must balance two considerations.
First, for low values of a given threshold, $\mu_*$,
the fraction of events with $\mu_\rel <\mu_*$
scales $\propto \mu_*^3$.  In particular, for bulge lenses, the
fraction of underlying events with low $\mu_\rel$ is
$f=(\mu_*/\sigma_\mu)^3/6\sqrt{\pi}\rightarrow
0.013(\mu_*/1.5\,\masyr)^3$, where $\sigma_\mu=2.9\,\masyr$ is the
dispersion of bulge stars, while the distribution for disk lenses is
qualitatively similar.  See Equation (22) and neighboring discussion
from \citet{masada}.  By construction, for $M=M_\odot$ contaminants,
$\mu_* = \mu_{\rm thresh}$, so at lower masses, $\mu_* =
(M/M_\odot)\mu_{\rm thresh}$.  Hence, setting $\mu_{\rm thresh}$ lower
very rapidly eliminates contaminants.

On the other hand, setting $\mu_{\rm thresh}$ lower also eliminates high
proper-motion BHs.  For example, if $\mu_{\rm thresh}$ were set at a lower
value, e.g., $\mu_{\rm thresh}=1\,\masyr$ to more aggressively screen against
contaminants, then this would also eliminate $10\,M_\odot$ BHs with
$\mu_\rel>10\,\masyr$, and it would eliminate $5\,M_\odot$ BHs with
$\mu_\rel>5\,\masyr$.  My purpose here is not to decide these issues but
rather to provide the mathematical framework for making these decisions
in the context of a detailed understanding of the available resources.

Finally, I conclude this section by addressing one practical issue.
The great majority of contaminants that survive
selection due to low proper motion will fail to yield interferometric
measurements of $\theta_\e$.  That is, adopting $\mu_{\rm thresh} =
1.5\,\masyr$, the Einstein radius of surviving contaminants will be
$\theta_\e < \mu_{\rm thresh}t_\e(M/M_\odot) = 1\,\mas
(t_\e/240\,{\rm day}) (M/M_\odot)$,
which is below the threshold of
detection by VLTI unless the event is extremely long.  Hence, unless
the observations are judged to be technically problematic, the issue
of a second observation will not arise.  Also, of course, even if
$\theta_\e$ is measured for such a long-event contaminant, its low
scientific interest would generally not justify additional
interferometric observations even if the four-fold degeneracy remained
unbroken.

\section{{$\bpi_\e$ Measurement Process for BHs}
\label{sec:bpie-measurement}}

The measurement of $\theta_\e$ from VLTI interferometry does not
require much more discussion: either the measurement can be made or it
cannot, with a sharp transition near $\theta_\e\sim 1\,\mas$,
depending somewhat on observing conditions and the availability of
reference stars.  If the measurement can be made, its fractional
precision will far exceed both the requirements of the experiment and
the fractional precision of the $\bpi_\e$ measurement.

By contrast, there are three strong reasons to undertake a comprehensive
analysis of the $\bpi_\e$ measurement.  First, the precision required
is substantially better than has previously been achieved, so it is
important to identify, at a theoretical level, the obstacles to high precision
measurements.  Second, the advent of VLTI interferometry measurements,
which were originally introduced to enable routine $\theta_\e$ measurements,
actually do provide new information that can greatly improve the precision
of the $\bpi_\e$ measurements.  Third, the $\bpi_\e$ measurement process
is relatively complex, so a review would be warranted even if there were
nothing new to report about it.

\subsection{{$\bpi_\e$ Precision Requirement}
\label{sec:require}}

As discussed in Section~\ref{sec:intro},  the $\theta_\e\ga 1\,\mas$ threshold
of VLTI measurements implies that $M=10\, M_\odot$ BHs will be accessible to
mass measurements provided that they have
$\pi_\e = \theta_\e/\kappa M \ga 0.012$.  Such BHs will have
$\pi_\rel = \theta_\e^2/\kappa M\ga 12\,\muas$, which corresponds to source-lens
relative distances down to
$D_{LS}\equiv D_S - D_L \simeq (\pi_\rel/\au)D_S^2 \sim 0.8\,\kpc$,
i.e., the regime of bulge lenses.  That is, at the $\theta_\e$ threshold, the
BH population is expected to be both plentiful and scientifically important.
Indeed, this threshold is just below the typical bulge-lensing relative
parallax, $\pi_\rel \sim 15\,\muas$,
where the population is likely to peak.  To obtain scientifically important
results for this population, the precision should be a factor of several
smaller than the threshold value, i.e., $\sigma(\pi_\e)\sim 0.003$.

Of course, if this precision cannot be achieved due to a combination of
economic and technological factors, then the experiment can still return
valuable information about the BH population, but at the outset we should
frame the problem in terms of achieving this goal.

\subsection{{$\bpi_\e$ Parameter Counting: 10, 9, 8}
\label{sec:parameter}}

The initial discussions of satellite parallaxes by \citet{refsdal66} and
\citet{gould94b} were framed in terms of proof of concept, and they
did not discuss measurement precision at all.  In these simplified treatments,
the Earth's orbital acceleration was ignored and the satellite was treated
as having a fixed offset from Earth.  The light curves from these two
observatories were treated as yielding the standard three parameters for
point lenses, i.e., $(t_0,u_0,t_\e)_{\rm ground}$ and $(t_0,u_0,t_\e)_{\rm sat}$,
and these quantities were combined,
as in Equation~(\ref{eqn:sat-parallax}).  In fact,
such ``Paczy\'nski fits'' require five parameters, with the other two
being the $f_S$, i.e., the source flux, and the blend flux $f_B$, which
does not participate in the event.  Thus, at this level, there appear
to be 10 parameters.  However, within the context of this simplified
rectilinear approximation, $t_{\e,\rm sat} = t_{\e,\rm ground}$, so there are
actually only 9 independent parameters.

This parameter reduction ($10\rightarrow 9$) is of fundamental importance
to the measurement process, in particular to the measurement of
$\Delta u_0= u_{0,\rm sat} - u_{0,\rm ground}$.  Writing the 5-parameter
microlensing equation explicitly,
\begin{equation}
  F(t) = f_S A(u[t]) + f_B; \qquad
  A(u) = {u^2 + 2\over u\sqrt{u^2+4}}; \qquad
  u(t) = \sqrt{u_0^2 + \biggl({t-t_0\over t_\e}\biggr)^2},
  \label{eqn:pacfit}
\end{equation}
we see that one of the partial derivatives $\partial F/\partial a_i$ with
respect to the 5 parameters $a_i$, namely $\partial F/\partial t_0$,
is odd in $(t-t_0)$, while the
remaining four are even in $(t-t_0)$.  Hence, assuming roughly uniform data
coverage, $t_0$ is essentially
uncorrelated with any other parameter, while the remaining four parameters
are correlated with each other.  Thus, from the standpoint of error analysis,
$\Delta t_0$ and $\Delta u_0$, which both appear in
Equation~(\ref{eqn:sat-parallax}), enter very differently.
That is, the error in $\Delta t_0$ is essentially just the quadrature sum
of the errors in $t_{0,\rm ground}$ and $t_{0,\rm sat}$, while the error in
$\Delta u_0$ is not a simple quadrature sum because $u_{0,\rm ground}$ and
$u_{0,\rm sat}$ are tied together via their correlations with their common
parameter, $t_\e$.  Furthermore, this common correlation potentially acts
in a very different way for the $\pi_{\e,\rm small}$ solutions (which
have the same signs for $u_0$), and the $\pi_{\e,\rm large}$ solutions (which
have opposite signs).

To elucidate this difference, I note that for the first ($\pi_{\e,\rm small}$)
case, Equation~(\ref{eqn:sat-parallax}) can be rewritten
\begin{equation}
  \bpi_\e = {\au\over D_\perp}{(\Delta t_0,\pm\Delta t_\eff)\over t_\e},
\label{eqn:sat-parallax2}
\end{equation}
where $t_\eff\equiv |u_0| t_\e$ is defined to be a strictly positive
quantity, and $\Delta t_\eff \equiv t_{\eff,\rm ground}-t_{\eff,\rm sat}$.
Then, the ``$\pm$'' in this equation distinguishes between the $(+,+)$
and $(-,-)$ cases.  In general, $u_0$ and $t_\e$ are anticorrelated, and
for high-magnification events, they are almost perfectly anticorrelated
because in this limit,
$A\rightarrow u^{-1}=1/u_0\sqrt{1 + [(t-t_0)^2/t_\eff^2]}$, so
$t_\eff\rightarrow {\rm FWHM}/\sqrt{12}$.  That is, because both $t_0$
(peak time of the light curve) and $t_\eff$ (scaled FWHM) are direct
observables, these have roughly comparable errors.

For the moment, I will specialize to an ideal case in which the ground
and satellite data sets are of similar quanitity and quality, and
I will consider the more realistic case that the ground data are
overall superior further below.

Then the uncertainty
due to the correlated parameters is almost entirely in the common denominator
of Equation~(\ref{eqn:sat-parallax2}), i.e., $t_\e$.  Furthermore, as a
practical matter, this error plays very little role, even if it is as high
as a few percent because it then induces the same few percent error in $\pi_\e$.

However, most events are not high-magnification, and for these, $u_0$ and $t_\e$
are far from perfectly anticorrelated.  Hence, for the generic case,
the $\Delta t_\eff$ errors
become substantially larger than the $\Delta t_0$ errors.
This problem is somewhat ameliorated by the fact that $t_\e$ is constrained
to be the same for both observatories, but the errors in the second component in
Equation~(\ref{eqn:sat-parallax2}), is still larger than the first.

The situaion is, in general, substantially worse for the $\pi_{\e,\rm large}$
case.  However, because $\pi_\e\ll 1$ for BHs, this case only occurs for
high-magnification events, $u_0\ll 1$.
Recall that for these, $u_0$ and $t_\e$ are highly anticorrelated, so that
$\Delta t_\eff = t_{\eff,\rm ground} + t_{\eff,\rm sat}$ has similar uncertainty
to $\Delta t_0$.  Hence, for the specific case of BHs, the errors for
$\pi_{\e,\rm large}$ solutions are not more problematic than for
$\pi_{\e,\rm small}$.

In fact, however, due to technological and economic contraints, the satellite
data stream will likely yield substantially larger errors to a simple five
parameter \citep{pac86} fit than the ground data.  Thus, the errors in
$|\Delta u_0|= u_{0,\rm ground} \pm u_{0,\rm sat}$ will be dominated by those
in $u_{0,\rm sat}$, and therefore, the correlations between
$u_{0,\rm ground}$ and  $u_{0,\rm sat}$ that are induced by the fact that
the two fits share a common $t_\e$ do not substantially ameliorate the problem..

Before continuing, I note that the idealization that led to
$t_{\e,\rm sat} = t_{\e,\rm ground}$, namely that
Earth and the satellite share a common rectilinear motion, may seem
excessively restrictive.  However, at the next level of approximation,
the two observatories each have rectilinear motion, but at different velocities.
As shown by \citet{gould95} this leads to a difference in $t_{\e,\rm sat}$ that is
completely specified by four quantities, i.e.,
$t_{\e,\rm ground}$, the particular solution's $\bpi_\e$, ${\bf D}_\perp$, and
$d{\bf D}_\perp/dt$, all of which are known.  Thus, the
role of the $t_\e$ constraint remains essentially the same, although it would
be more cumbersome to express it explicitly.  Finally, for the general case
of full orbital motion of Earth and the satellite, the same information
content is preserved in the standard nine-parameter characterization,
$(t_0,u_0,t_\e,\bpi_\e,f_{S,\rm ground},f_{B,\rm ground},f_{S,\rm sat},f_{B,\rm sat})$,
where $\bpi_\e$ is often expressed in equatorial coordinates,
$\bpi_\e = (\pi_{\e,N},\pi_{\e,E})$.  In this case, $(t_0,u_0,t_\e)$ are the
\citet{pac86} parameters in the geocentric frame \citep{gould04}.  Such
fits preserve the constraint that the event has the same $t_\e$ in the
heliocentric frame, while it simultaneously and automatically
incorporates information from the ground-based (annual) parallax
that was discussed in Section~\ref{sec:intro}.  In fact, it is now customary to
compare three fits as a check on systematics in either the ground-based
or satellite data.  The first is the nine-parameter fit just described.
The second is a ground-only fit, i.e., ignoring the satellite data.
The third is a satellite-``only'' fit \citep{kb180029}, which mimics the
common-$t_\e$ fit that was described near the beginning of this section.
In this case, one fits the satellite data, but with $(t_0,u_0,t_\e)_{\rm ground}$
fixed at the fit to the ground data and sets
$t_{\e,\rm sat} = t_{\e,\rm ground}$.

In the paper in which I first analyzed the errors in satellite parallax
measurements \citep{gould95}, I adopted the
assumption (now clearly unreasonable, see below) that the ground and
satellite photometry would be of similar quantity and quality.  I also did not
impose a constraint on $t_\e$, such as $t_{\e,\rm sat} = t_{\e,\rm ground}$,
in order to demonstrate that the independent fits to the two data sets
would yield different values of $t_\e$, from which one could determine which
of the four $\bpi_\e$ solutions was correct.  I was alarmed to find that,
within this formalism, $\Delta u_0$ had huge errors.  As I have just
described in this section, if I had imposed a constraint on
$\Delta t_\e= t_{\e,\rm ground} - t_{\e,\rm sat}$ (which would be different for
each for the four solutions), then the errors in $\Delta u_0$ would have
been much smaller.

I therefore sought a different constraint: I realized that if the filter+CCD
response were exactly the same for the satellite and ground observatories,
then one would know, a priori, that $f_{S,\rm sat}=f_{S,\rm ground}$.  Without
going through all the details, one can see immediately that such a constraint
would play essentially the same role as the $t_{\e,\rm sat}=t_{\e,\rm ground}$
constraint described above.  For example, in the limit of high magnifcation,
$f_S/u_0=A_\max$, while $A_\max -1 = F_{\rm peak}-F_{\rm base}$ is a direct
observable, and so has small errors that, in particular, are uncorrelated
with other parameters.  Here, $F_{\rm peak}$ and $F_{\rm base}$ are the
observed flux at the peak and baseline of the event, respectively.
That is, for high-magnification events, $f_S$ and $u_0$ are almost perfectly
correlated.  Just as for the $t_\e$ constraint, the correlation is weaker for
low magnification events, but
it remains significant.  Based on this reasoning, I concluded that such
a flux constraint was an essential condition for a parallax satellite.
In particular, when Michael Werner (1998, private communication) contacted
me a few years later about using the prospective {\it SIRTF}
(later {\it Spitzer}) infrared mission as a parallax satellite, I told him
flatly that this was impossible due to the vastly different wavelengths
($3.6\,\mu$m versus $0.8\,\mu$m) of the space and ground data.  After Werner
persisted, I developed an approach (valid only for relatively long events)
that ignored the ``corrupted'' $\Delta u_0$ parallax component and used only the
robust $\Delta t_0$ component.  This 1-dimensionsl (1-D) parallax measurement
would be combined with independent 1-D
parallax information from the ground-based data,
which can be extracted from moderately long events \citep{gmb94}.
This led to a theoretical paper \citep{gould99} and ultimately to the first
space-based $\bpi_\e$ measurement \citep{dong07}.

Although the argument that I gave for the necessity of a flux constraint
\citep{gould95} was incorrect, the conclusion was valid.  That is, under
the assumption of that paper of equal-quality space and ground data,
a flux constraint would not substantially improve the $\bpi_\e$ precision
relative to the $t_\e$ constraint, which is automatically incorporated
into any real fit.  However, when the satellite data are substantially
inferior in quantity and/or quality (as will often be the case),
then (as I have just described above) the $t_\e$ constraint does not
by itself improve the precision of the $\bpi_\e$ measurement.

This tangle of issues, seemingly of only historical interest, are all
directly relevant to the present problem of BH mass measurements.

The first point is that it is certainly cost-ineffective, and perhaps
not even technologically feasible for a space observatory to conduct a
high-cadence survey over the tens of square degrees that are covered
from the ground to locate BH candidate events.  Thus, the space
telescope must be sequentially pointed at candidates that are alerted
from the ground.  If there are, say, 50 events that, at a given time,
are either BH candidates or could plausibly develop into BH
candidates, and if the slew plus exposure time is 20 min, then the cadence
will be $\Gamma=0.06\,{\rm hr}^{-1}$ compared to $\Gamma=1\,{\rm hr}^{-1}$ from
the ground.  While this will be partially compensated by
higher-quality space conditions, the ground data will still be
significantly better.  In addition, because events are only alerted
after they have noticeable deviations from baseline, the initial rise
will be lost from space (but not from the ground).
The satellite should be separated by of order $1\,\au$. If it is in an
Earth-trailing orbit, then for events whose rise takes place in the Northern
spring, $D_\perp$ will be small, greatly reducing the value of the observations.
If the satellite is in an Earth-leading orbit, then the same would hold
for events whose fall takes place in the summer.  Therefore, unless
extraordinary resources are applied to the space component of the project,
the satellite data will be inferior.  Hence, a flux constraint is certainly
needed.

Second, while the huge {\it Spitzer}-ground wavelength ratio was not the
show-stopper that I imagined when I was approached by Mike Werner, and neither
did it restrict {\it Spitzer} parallaxes to a narrow sub-class of events
(as I imagined in \citealt{gould99}), it did end up degrading the parallax
measurements to below the quality that will be required for BH mass
measurements.  That is, it was subsequently found that fluxes in different
bands could be brought to the same scale via color-color relations
\citep{gould10b,mb11293}, and it subsequently became routine to apply this
technique to tie the ground and {\it Spitzer} flux scales together
despite the huge difference in wavelengths \citep{yee15,calchi21}.  However,
due in part to the very different wavelengths, it generally proved possible
to do this only up to a precision of a few percent, which then induces errors
in $u_{0,\rm sat}$ at the same level.  Hence, for typical cases of BH events,
i.e., $u_0\sim 0.3$, the errors in $\pi_\e$ would be of order 0.01, i.e., of
the same order as the value of $\pi_\e$ for bulge BHs.

Thus, to reduce this problem to the absolute minimum, the filter+CCD
response should be as similar as possible for the specific application
of the small-$\pi_\e$ measurements that are needed for BHs.  Because
of atmospheric absorption, it is probably not possible to have exactly
the same response.  However, the responses can be made very similar, which
will then allow color-color-based corrections of the remaining small
difference.  This is one of the major advantages of using a satellite that
has been built specifically for microlensing parallaxes, as opposed to
a a general-purpose satellite like {\it Spitzer}.

Third, even though (contrary to \citealt{gould99}) directional
information is not absolutely essential for satellite-based parallax
measurements in general, it will be crucial for some BH measurements
and very helpful for most of the rest.  That is, in some cases, it
will not be possible to fully break the \citet{refsdal66} four-fold
degeneracy based only on the light curves from the satellite and
ground, so having directional information will be crucial.  However,
even when the correct solution is identified, the direction will be
much more precisely determined from VLTI than from the light curve,
which implies that the combined measurement will be improved by
incorporating the VLTI direction.  For cases that the light-curve
based $\bpi_\e$ errors are isotropic, this improved direction will
yield little or no improvement in the parallax amplitude,
$\pi_\e=|\bpi_\e|$ (which is what is needed for the mass measurement).
However, for the majority of cases that the errors in $\Delta u_0$ are
substantially larger than those in $\Delta t_0/t_\e$, precise external
information on the direction will significantly improve the precision
of $\pi_\e$.

If the flux constraint were exact, then the number of free parameters
of the fit would be reduced from 9 to 8.  In actual fits, the flux
constraint is incorporated as a $\chi^2$ penalty on the source-flux ratio,
and so, formally, there are still 9 free parameters.  However, in the limit
of small error bars for this ratio, the final results from this 9-parameter
fit are essentially the same as from imposing a fixed flux ratio (8 parameters).
I will show below that ``realistic'' error bars on the flux ratio yield
results that are close to this limit.  Hence, the fits can be thought
of qualitatively as having 8 free parameters, i.e.,
$(t_0,u_0,t_\e,f_S,f_B)_{\rm ground}$, $f_{B,{\rm sat}}$, and $(\pi_{\e,N},\pi_{\e,E})$.

Finally, there is one further technical issue that is buried deep in
the analysis of \citet{gould95}, but which I have, for simplicity of
exposition, avoided up to this point: blending. In microlensing, the source
flux, $f_S$, is not a direct observable: only $f_{\rm base}=f_S + f_B$
can be directly inferred from the light curve.  As mentioned above,
$f_S$ is entangled with three other parameters, and the precision of the
$f_S$ determination crucially depends on the wings of the light curve.

This can be a particularly serious issue for high-magnification events
of very faint sources.  Then, the wing data near $u\sim {\cal O}(1)$
that are needed to determine the correlated parameters
$(u_0,t_\e,f_S,f_B)$ can be too noisy for a precise measurement, and
(what is more troubling), systematic errors induced by long-term
trends, whether of astrophysical or instrumental origin, in the
baseline flux can degrade the accuracy even further.  In such a case,
the error in $t_\e$ can be several tens of percent, even though
$\Delta t_0$ and $\Delta t_\eff$ are measured very accurately from the
data taken near peak.  A good example is given by \citet{mb11293}, but
the situation can be substantially worse for heavily extincted events,
for which $f_S$ can be extraordinarily faint in the optical, while the
highly-magnified $K$-band flux permits an excellent VLTI measurement
of $\theta_\e$.  In such a case, we see from
Equation~(\ref{eqn:sat-parallax2}) that the fractional error in
$\pi_\e$ is equal to the fractional error in $t_\e$.

For such cases, the VLTI measurement automatically yields additional
information that constrains $t_\e$, namely the flux ratio, $\eta$, of the
minor image relative to the major image.  The individual magnifications
of the images are given by $A_\pm = (A\pm 1)/2$, which implies
$\eta=(A-1)/(A+1)$.  Using $u^2=2[(1-1/A^2)^{-1/2}-1]$, and after some algebra,
one finds
\begin{equation}
  u = \eta^{-1/4} - \eta^{1/4}.
  \label{eqn:u-r-relation}
\end{equation}
In particular, for $A\gg 1$, $u\simeq (1-\eta)/2$,
so that $\sigma(u) \simeq \sigma(\eta)/2$.  To be specific, consider an
$A_\max=100$ event for which a VLTI measurement is made (as would be typical)
at $t = t_0+t_\eff$, a time that would be known very precisely because
both $t_0$ and $t_\eff$ are precisely measured.  Then $u_0 = u/\sqrt{2}$,
so $\sigma(t_\e)/t_\e\simeq \sigma(u_0)/u_0 \simeq A_\max \sigma (\eta)/\sqrt{8}
\rightarrow 35\,\sigma(\eta)$.

In the first interferometric measurement of a microlensing event,
\citet{kojima1} found $\sigma(\eta) = 0.032$, which would not appear to
be very promising for this example.  However, this ``poor'' precision
was actually due to the fact that the measurement was made under
marginal conditions.  In the more recent case of KMT-2023-BLG-0025,
Subo Dong (2023, private communication) found $\eta=0.26133\pm
0.00045$ and $\eta=0.20901\pm 0.00081$ at two observational epochs.
In the above example, these $\sigma(\eta)$ would yield
$\sigma(t_\e)/t_\e = 1.6\%$ and $2.8\%$, respectively.

Thus, in principle, interferometric flux-ratio measurements can play
an important role in at least some cases.  However, it would be
premature to systematically assess this role at the present time.
First, it will be necessary to test whether the very high formal
precision being reported is confirmed by independent tests of the
accuracy of these measurements.  Because essentially every successful
interferometric observation will yield a flux-ratio measurement and
the formal errors for many of these will be of the same order as, or
larger than, those of the flux ratios predicted from the fit to the
light curve, it will be possible to test the accuracy of these
measurements as a bi-product of the black-hole mass-measurement
program.  For the moment, it is only necessary to be aware that flux
ratios can play an important role in some cases.

\section{Examples}
\label{sec:examples}

In order illustrate and possibly refine the analytic arguments
given above, I present detailed analyses of two simulated events.

\subsection{Simulation Characteristics}
\label{sec:simchars}

For the ground-based data, I adopt the KMTNet telescope and instrument
characteristics, but with a different observing strategy, i.e., all
fields would be monitored with a cadence of $\Gamma=1\,{\rm hr}^{-1}$
as opposed to the current multiple-cadence strategy that is described by
\citet{eventfinder}.  This would allow for about $100\,{\rm deg}^2$ to
be covered at this cadence.  Specifically, $10^4$ photons are collected
in a 1-minute exposure at $I=18$.  I assume that observations can be
carried out beginning at nautical twilight provided that the target is
at least $35^\circ$ from
the horizon.  I assume that at the three observatories (CTIO, SAAO, SSO),
the time lost to weather and other problems is (15\%, 25\%, 30\%).
Because the evolution of the event is slow compared to the diurnal cycle, and
even to weather cycles, I bin the observations by day, taking account of the
average observing efficiency mentioned above.  For the extreme wings of the
season, when there is an expectation of less than 1 observation per day,
I inflate the errors accordingly, so as to simulate a fractional observation.
I assume a mean background flux equivalent to an $I=16.5$ mag star, i.e.,
$4\times 10^4$ counts for a 1 minute exposure.

For the satellite, I choose a $0.5$ meter mirror, with similar throughput
to the KMT telescopes, and with 50 targets that must be observed each
observing cycle.  I assume 19 minutes per target, for a net cadence of
$\Gamma=1.5\,{\rm day}^{-1}$.  I assume that 16 minutes can be used
for (possibly stacked) exposures, with the remaining 3 minutes used
for readout and slewing.  Note that the satellite field of view can be small,
so readout can be very rapid.

For the satellite orbit, I adopt the {\it Spitzer} orbit from 2014, when
it was trailing the Earth, very close to the ecliptic, with a distance
$D_{\rm sat}\simeq 1.3\,\au$.  I adopt a Sun-exclusion angle of $45^\circ$.
In order to facilitate independent investigations of my results, I quote
Heliocentric Julian Dates (HJD) from 2014.  Of course, one could equally
well add $(N - 2014)*365.25$ to these dates, where $N$ is the year of
the anticipated experiment.  For the satellite, I assume background light
(mainly due to ambient stars) equivalent to an $I=19$ star.

I require three conditions for satellite observations to take place.
First, the target must be beyond the $45^\circ$ Sun-exclusion angle.
Second, the source must have already entered the Einstein ring, as
seen from the ground (although observations can continue after it
has left the Einstein ring).  Third, at least 30 days must have
elapsed since the first observation of the season (in early February).
The latter two conditions are to allow information to accumulate that
enable decisions on what events are plausible long-event candidates.
However, I should note that with the satellite trailing Earth at
$D_{\rm sat}\sim 1.3\,\au$
(as I am assuming for these examples), the third condition is redundant:
any observation satisfying the first two conditions will also satisfy the third.
In particular, for the event coordinates that are specified below,
no satellite observations can take place prior to 23 April, while for
other bulge locations the starting date differs by just a few days.
Thus, even if the satellite trailed by just $D_{\rm sat}\sim 0.6\,\au$,
a possibility that I will discuss in Section~\ref{sec:discuss},
the third condition
would still be redundant because the Sun-exclusion angle would be satisfied
only about 47 days earlier, i.e., on about 7 March.  That is, for a nearly
circular Earth-like orbit, the time lapse of the satellite position relative
to Earth is given by $\Delta t = 2\sin^{-1}(D_{\rm sat}/2\au)({\rm yr}/2\pi)$,
or 82 days and 35 days in the two cases.

For both simulated events, I adopt $I_S=19$ for the source flux, and I assume
$M=12\,M_\odot$, $\pi_\rel=12.7\,\muas$, $\mu_\rel=5.87\,\masyr$, and
$u_{0,\rm ground}=+0.3$.  These values lead, in both cases to
$\theta_\e=1.14\,\mas$, $\pi_\e=0.0114$, and $t_\e= 69\,{\rm day}$.
For Event~1, I adopt $t_0=6779$ (i.e., 1 May) and
$\bpi_\e=(\pi_{\e,N},\pi_{\e,E})=(+0.0081,+0.0080)$, while
for Event~2, I adopt $t_0=6902$ (i.e., 1 September) and
$(\pi_{\e,N},\pi_{\e,E})=(+0.0081,-0.0080)$.  That is, Event~1 peaks relatively
near the beginning of the microlensing season and has a proper motion
in the north-east quadrant, while Event~2 peaks relatively
near the end of the microlensing season and has a proper motion
in the north-west quadrant.  Here, dates are expressed as
HJD$^\prime={\rm HJD}-2450000$.  Finally, both events have the same
coordinates: $(\alpha,\delta)_{\rm J2000} =$ 17:53:00 $-$29:00:00, i.e.,
near the peak of the observed surface density of microlensing events.

Overall, there are two broad questions that must be addressed.
First, can the event be sufficiently well understood in time to make an
informed decision as to whether to undertake a VLTI interferometric measurement?
Second, do the ensemble of observations (including the interferometric
measurement) result in well-determined microlens parallax $\bpi_\e$
(and hence, mass, distance, and transverse velocity measurements)?

Logically, the first question should come first.  However, this issue
can be handled much more flexibly than the second because if there
is not sufficient information at a given date, the decision can often
be postponed until there is more information.  The cost is that the
event will be fainter (and in some cases closer to the horizon), which
generally make the interferometric measurement more difficult.  Whether
the measurement then remains feasible depends on many details, such as
the $K$-band brightness and the availability of reference stars.
In order to simplify the discussion, I just assume that the interferometric
measurement will be made at $t=t_0 + t_\eff$.  This allows me focus on the
second, more fundamental, question, i.e., how well can $\bpi_\e$ be measured.
I then treat the question of how well the event is understood at
$t=t_0 + t_\eff$ in this context.

Figure~\ref{fig:lc} shows the simulated data for these two events in
main panel.  The decision times ($t=t_0 + t_\eff$) for the interferometric
observations are indicated.  The north and east components of the
Earth-satellite separation ${\bf D}_{\rm sat}$ are shown as a function
of time in the upper panel.

\subsection{Event~1}
\label{sec:event1}

\subsubsection{Ground-Only Analysis}
\label{sec:ground1}

I begin with a ground-only analysis for two reasons.  First, we would
like to know whether a ground-only lightcurve (plus interferometry)
can yield a viable $\bpi_\e$ measurement, and if not, how far it falls
short with respect to this objective.  Second, we want to understand
explicitly, what role the ground-only data play in resolving the four-fold
degeneracy that was identified by \citet{refsdal66}.

The right two panels of Figure~\ref{fig:ground1} show the final parallax
measurement based on ground-only data, under the respective assumptions
that $u_0 > 0$ and $u_0 < 0$, which is not known a priori.  By themselves,
the ground data clearly provide no information on the magnitude of $\pi_\e$,
and hence on whether the lens is a BH or not.  However, these data can
potentially limit or rule out the possibility that the ``alternative''
($\pi_{\e,\rm large}$) solution (from the \citealt{refsdal66} analysis) is
correct.
Unless this solution lay on or near the one-dimensional structures shown
in Figure~\ref{fig:ground1}, there could be no local minimum in the $\chi^2$
surface at the ``alternative'' solutions.  Of course, one could determine
whether a local minimum existed by conducting a grid search over the
$\bpi_\e$ plane, but the point of the present exercise is to understand
why we would expect, or not expect, such a minimum to exist.  I will
return to this issue in Section~\ref{sec:ground2}.

The reason for the one-dimensional structures in the two right-hand
panels is well understood, at least in the limit of events with short effective
timescales, $t_\eff\ll {\rm yr}/2\pi$, i.e., the same limit that underlies the
parallax-satellite analyses of \citet{refsdal66} and \citet{gould94b}.
For short $t_\eff$ events, Earth's (projected) acceleration toward the 
position of the Sun, {\bf S}, is approximately constant,
which induces an asymmetry in the light curve with respect to $t_0$,
yielding a measurement of the component $\pi_{\e,\parallel}$, whose
direction is therefore
\begin{equation}
  \label{eqn:psipar}
  \Psi_\parallel = \tan^{-1}{-S_E(t_0)\over -S_N(t_0)}.
\end{equation}
The poorly measured component is then $\Psi_\perp = \Psi_\parallel + \pi/2
= \tan^{-1}([-S_N(t_0)]/S_E(t_0))$.  See Figure~3 of \citet{mb03037} for sign
conventions and for an example of a highly linear structure for an event
with $t_\eff=2\,$day (keeping in mind that MOA-2003-BLG-37 peaked in the
summer, whereas Event~1 peaks in the spring).

For Event~1, $(S_N,S_E)(t_0)=(-0.075,+0.742)\,\au$.
Hence, the expected direction of the linear structures in
Figure~\ref{fig:ground1} is $\Psi_\perp=5.8^\circ$,
north through east.

While many such structures have been noted in the literature for short
$t_\eff$ events, they have not, to the best of my knowledge, been
systematically studied for moderately long events like Events 1 and 2,
which have $t_\eff=21\,$day.
From Figure~\ref{fig:ground1}, we see
that the actual contours deviate from the short-$t_\eff$ prediction (and
past short-$t_\eff$ experience) in two key ways.  First, both sets of contours
are curved, and with very similar curvature.  Second, neither of the tangents
at the best-fit values (i.e., centers of the structures) matches the
$\Psi_\perp = 5.8^\circ$ expectation.  Instead, the $u_0>0$ tangent
is $\sim 12^\circ$, while the $u_0<0$ tangent is $\sim 0^\circ$.  Hence, they
agree with the expectation in their average, but not individually.

Nevertheless, regardless of whether these 1-D structures exactly
match theoretical expectations derived from the short-$t_\eff$ analysis,
they greatly reduce the chances that the $\pi_{\e,\rm large}$ solutions
will survive the joint fit to the Earth and satellite data.  In fact,
as I will show in Section~\ref{sec:ground2}, the elimination of the
$\pi_{\e,\rm large}$ solutions is not really a matter of ``chance''
(as might appear from the present discussion).  For now, I just note that
for events that, like Events 1 and 2, lie toward the bulge and
several degrees south of the ecliptic, $S_N$ retains a small negative
value through most of the microlensing season, while $S_E$ moves from
large positive values to large negative values.  Thus $\Psi_\perp$
rotates counterclockwise over the course of the microlensing season
from slightly east of north to slightly east of south.

The left panel of Figure~\ref{fig:ground1} shows the parallax
measurement derived from ground data at the time that I have specified
for a ``decision'' regarding interferometric observations, i.e., at
$t=t_0+t_\eff$.  See also Figure~\ref{fig:lc}.  The first point is
that the ground-only data do not contribute significant information
toward making this decision compared to the information that will be
available from the combined ground and satellite observations, which I
will discuss in Section~\ref{sec:interferometry1}.  Thus the panel may
appear to be of little interest.

However, it is also of some value to consider how one would react to
such a measurement if (as is true at present) there were no satellite,
and hence no satellite data.  First, one would conclude that $\pi_{\e,\parallel}$
(the only measurable component) was consistent with zero, and so (given
the long timescale, $t_\e=69\,$day) that this was a plausible BH candidate.
From there, one would have been led to ask: under what conditions could
an interferometric measurement yield a BH ``detection'', and if so, then
also a reliable
mass measurement.  One would consider various possible values of $\pi_\e$
that are consistent with the left panel, and simulate ground data from
the rest of season.  For example, for a trial $\pi_\e=0.0114$, such
simulations would yield $\bpi_\e$ maps similar to the two right panels.
As I will show just below, one would then immediately conclude that
the parallax error (even with the direction of $\bpi_\e$ determined
from interferometry), would be of order $\sigma(\pi_\e)\sim 0.015$, so
that a $3\,\sigma$ measurement would be possible only if $\pi_\e\ga 0.05$
i.e., $\pi_\rel = \kappa M \pi_\e^2 \ga 0.2\,\mas$ (or $0.1\,\mas$), and
$\mu_\rel = \kappa M \pi_\e/t_\e \ga 22\,\masyr$ (or $11\,\masyr$) for an
$M = 10\,M_\odot$ (or $5\,M_\odot$) BH.  While the first (higher-mass) scenario
is quite unlikely, the lower-mass scenario is not implausible, and
hence it is likely that the interferometric observations would have
been undertaken.

We can judge the results from Figure~\ref{fig:ground1small}, which provides
zooms of the two right panels of Figure~\ref{fig:ground1}.  The blue rays
show the directions of $\bpi_\e$ implied by the interferometric measurement
under the assumptions of $u_0>0$ and $u_0<0$, respectively.  That is, because
$\Phi_\pi = \tan^{-1}(\pi_{\e,E}/\pi_{\e,N}) = 44.6^\circ$, and
$\phi = \cot^{-1}(\delta t/t_\eff)\rightarrow 45^\circ$, while $u_0>0$,
the major image will lie at $\psi = \Phi_\pi + \phi + \pi= 269.6^\circ$
relative to the $y$-axis, i.e., approximately due west.  Then, of course,
if one interprets this measurement under the (correct) assumption that
$u_0>0$, one will recover $\Phi_\pi = \Psi - \phi - \pi= 44.6^\circ$.
However, under the assumption that $u_0<0$, one will infer
$\Phi_\pi = \Psi + \phi - \pi= 134.6^\circ$.  More generally, the
$u_0<0$ solution will always lie $2\phi$ counterclockwise from the
$u_0>0$ solution and, in particular, $90^\circ$ counterclockwise
if the observation is
at $t=t_0+t_\eff$.  See Equations~(\ref{eqn:phidef}) and (\ref{eqn:Phidef}).

The net result is that one would conclude that $\pi_\e<0.05$ at the $3\,\sigma$
level, but with no lower limit (even at $1\,\sigma$), and one would
simultaneously obtain a relatively precise measurement of
$\theta_\e=1.11\,\mas$.  Together these would imply
$M=\theta_\e/\kappa\pi_\e > 2.7\,M_\odot$ and
$\pi_\rel = \theta_\e\pi_\e < 57\,\muas$.  This would be a very secure
detection of either a BH or neutron star (NS), with the BH strongly
favored, partly by the fact that such massive NSs are expected to be
rare, and partly by the fact that such low values of the measured mass
are permitted only at $3\,\sigma$.  However, there would be no
information at all on the mass of the BH, and there would be no
clear indication of whether it was in the disk or the bulge.

\subsubsection{Ground+Satellite Analysis}
\label{sec:sat1}

Figure~\ref{fig:2comb1} shows the results of simultaneously fitting
the ground and satellite data under three different assumptions about
the flux constraint, i.e., that the ratio of source fluxes as seen from
the ground and the satellite can be constrained to
2.5\% (lower panels),
0.25\% (middle panels), and
0.025\% (upper panels).  The first is essentially no constraint because the
source fluxes from the fit can be determined to better than this precision.
I consider the second to be realistic, provided that the wavelength dependence
of the satellite and ground systems are similar.  I do not consider the
third to be realistic, but it represents a theoretical limit on
what can be achieved by externally constraining the flux ratio.
The $\chi^2$ contours are shown relative to the minimum in each panel.
The $\chi^2$ difference between the two panels in the same row
are indicated within the $u_0<0$ panels.

The first point to note is that the improvements from the lower 
to middle panels is substantially larger than than that from the
middle to the upper panels.  Similarly, the improvements in $\Delta\chi^2$
also show a much larger jump.
This implies that the ``realistic'' flux constraint is both
useful in constraining $\bpi_\e$ and cannot be much improved upon, even
by exceptional efforts.  Henceforth, I will therefore focus on the middle
panels.

\begin{figure}
\plotone{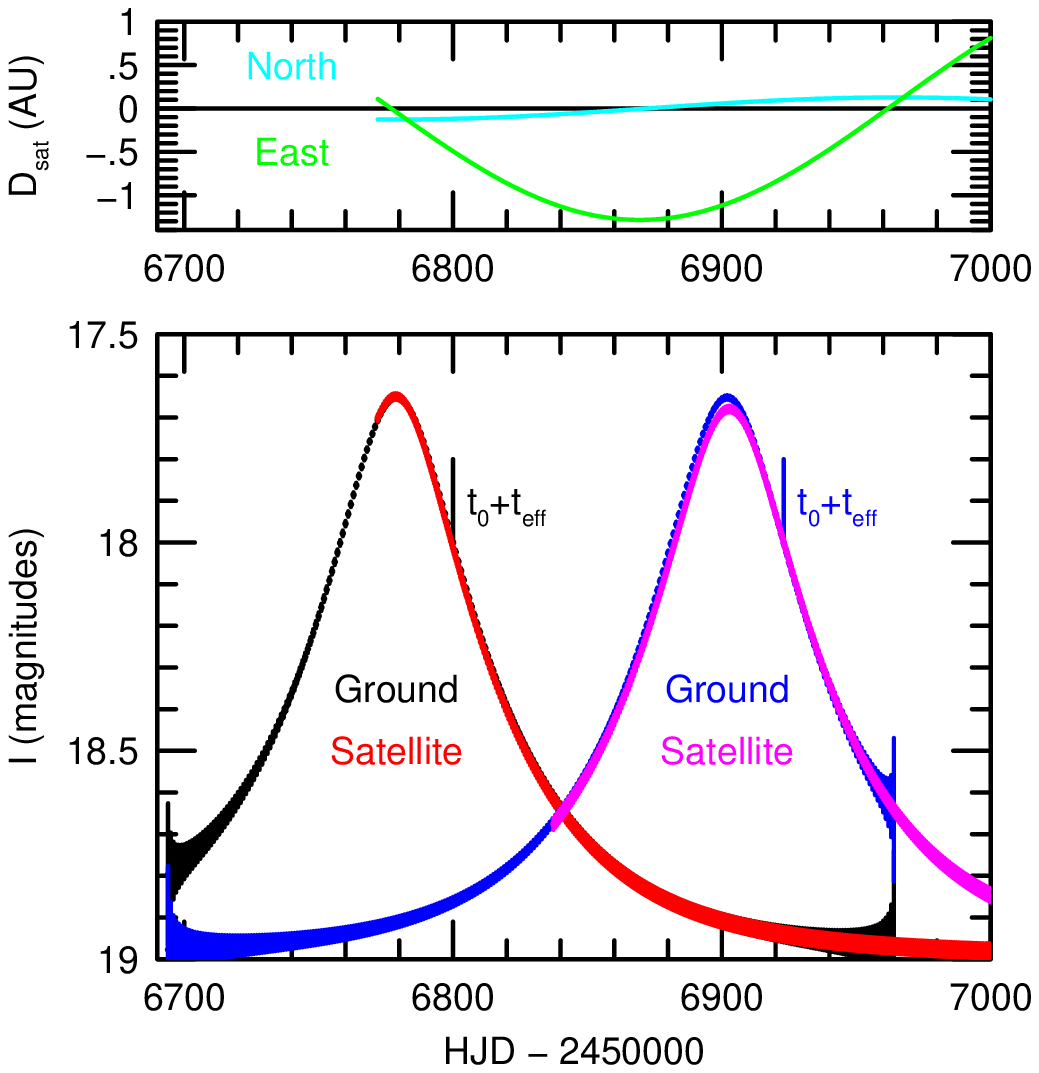}
\caption{Light curves for simulated Events 1 (left) and 2 (right).
The ground data are shown, respectively, in black and blue,
while the satellite data are shown in red and magenta.  The ground
data are binned by day, based on the mean number of observations expected
as a function of time of year.  The north and east components of the
Earth-satellite distance ${\bf D}_{\rm sat}$ are shown in the upper panel
for the epochs when data can be taken.  That is, the epochs when the satellite
would be pointed $<45^\circ$ from the Sun are excluded.
}
\label{fig:lc}
\end{figure}

\begin{figure}
\plotone{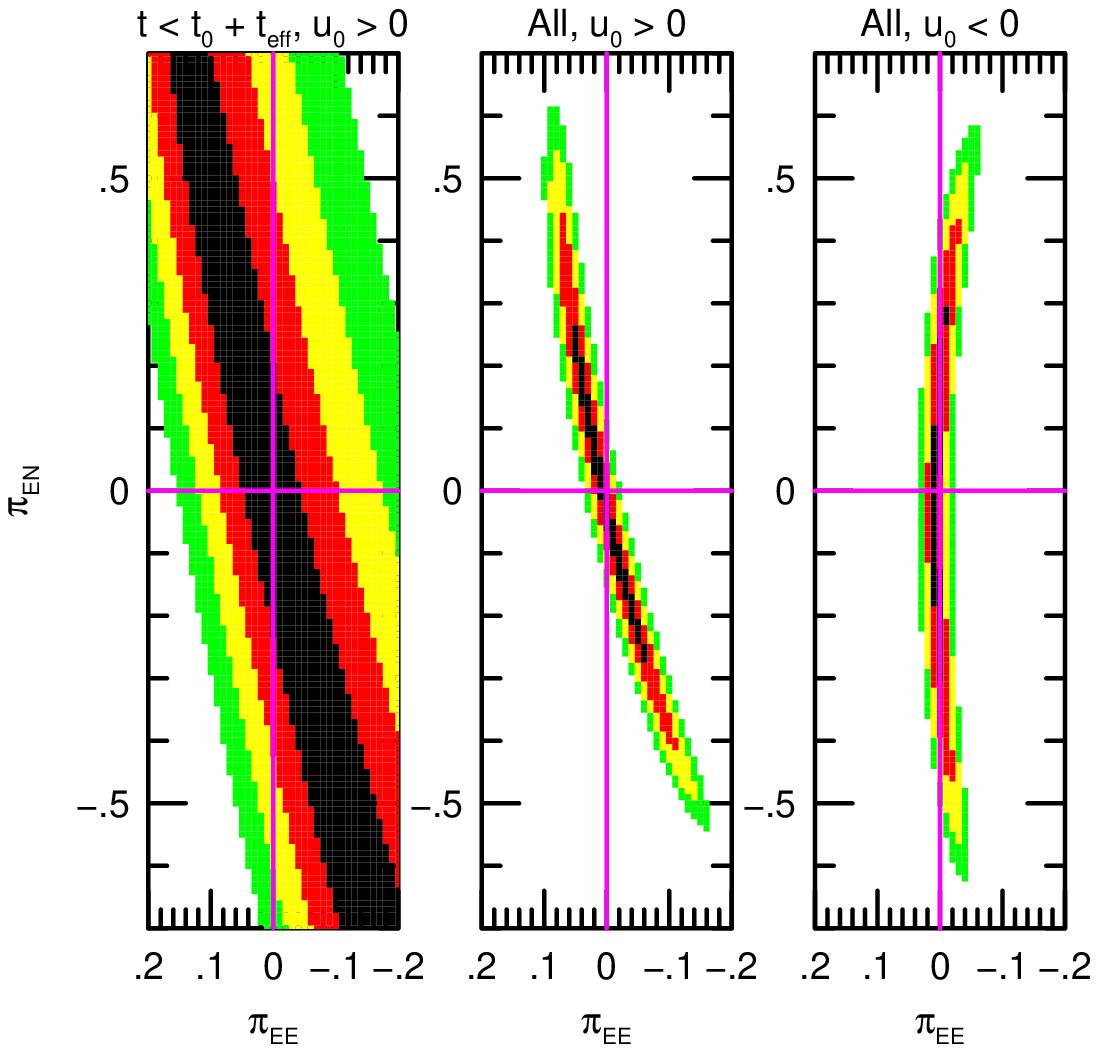}
\caption{Ground-only $\bpi_\e$ contours for Event~1.  In this, and all
  subsequent figures, the contour steps are set at
  $\Delta\chi^2 = 1, 4, \ldots, n^2$.
}
\label{fig:ground1}
\end{figure}

\begin{figure}
\plotone{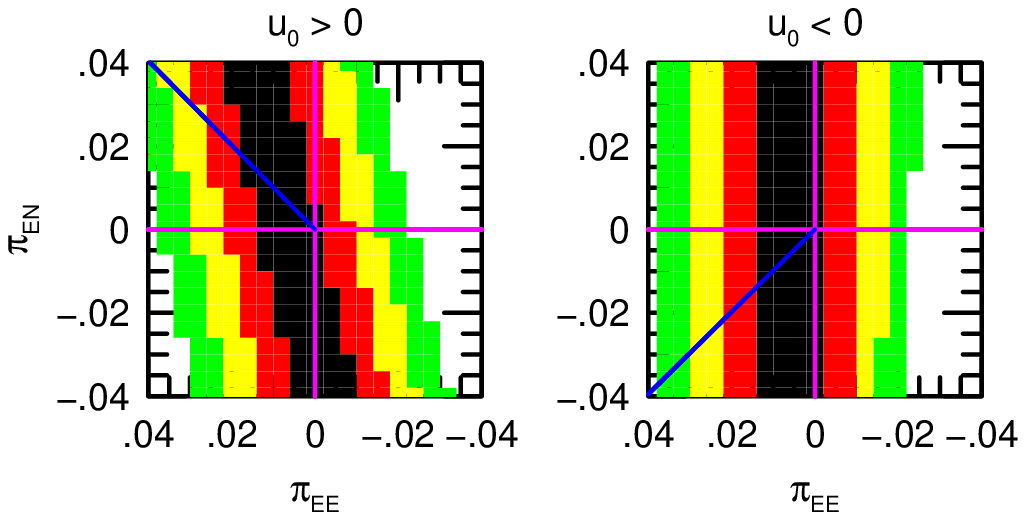}
\caption{Zoom of the two right panels from Figure~\ref{fig:ground1}.
  The blue rays in this figure (and all subsequent figures) represent the
  directions inferred from the interferometric measurement.
}
\label{fig:ground1small}
\end{figure}

\begin{figure}
\plotone{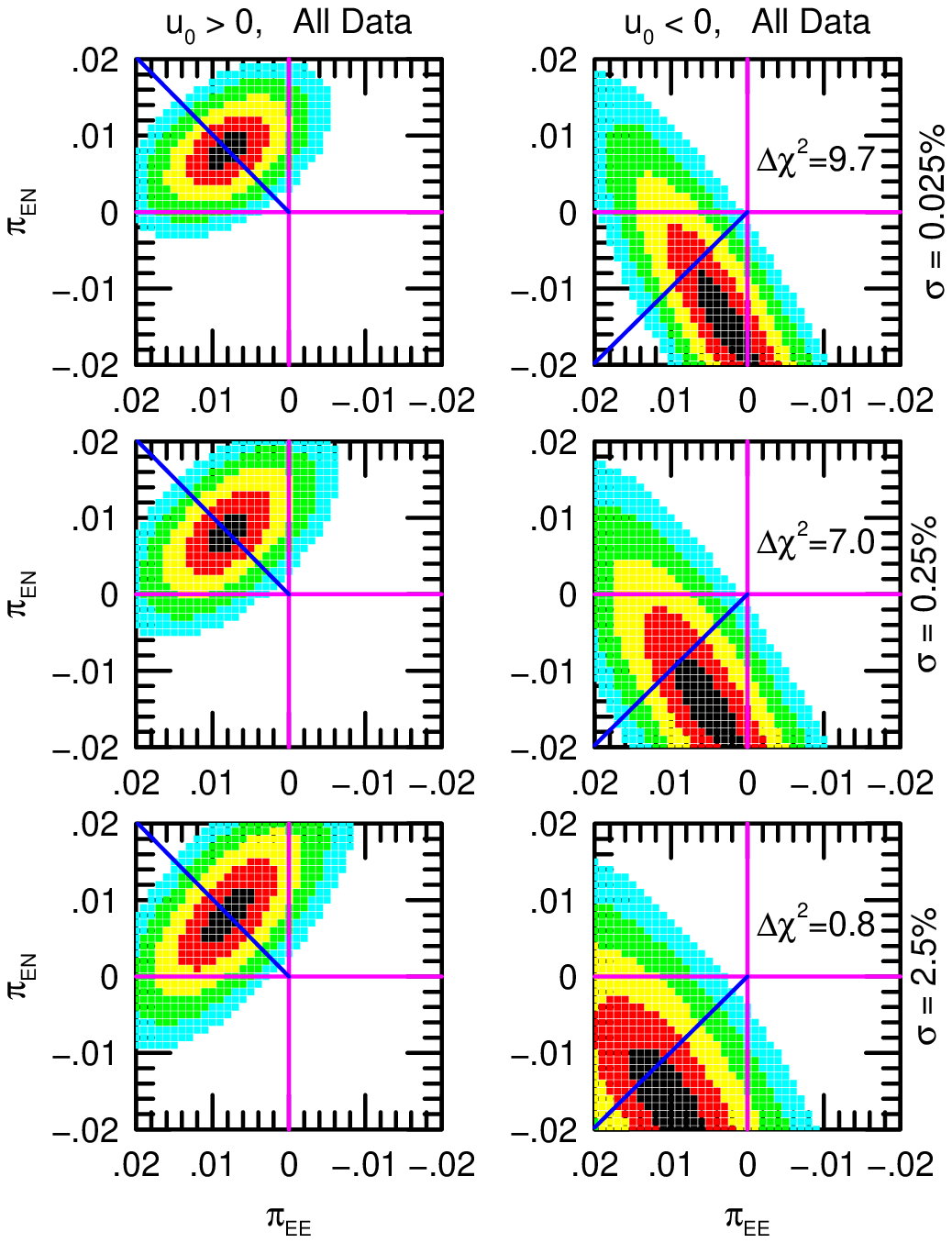}
\caption{$\bpi_\e$ contours from combined ground and satellite data for Event~1
  for three different values of the constraint on the ratio of source
  fluxes as seen from the ground and the satellite, which are indicated
  to the right.  Also shown is the $\Delta\chi^2$ difference between the
  minima of the $u_0<0$ solution relative to the $u_0>0$ solution.
}
\label{fig:2comb1}
\end{figure}

Second, for both $u_0>0$ and $u_0<0$, the contours are elongated at a
substantial angle relative to the cardinal axes.  This is somewhat
surprising in light of the discussion in Section~\ref{sec:parameter},
wherein I argued that the errors in the $\Delta t_0$ and $\Delta
t_\eff$ should be uncorrelated, which would imply that any departure
from isotropy in the error ellipses should take the form of elongation
either parallel or perpendicular to ${\bf D}_\perp$.  While it is true
that ${\bf D}_\perp$ changes substantially over the course of the event,
its direction is remains roughly fixed along the east-west axis.  See
the upper panel in Figure~\ref{fig:lc}.  In particular,
\citet{gould95} had argued that the errors in $\Delta u_0$ should be
larger than those in $\Delta t_0/t_\e$, which would lead to a
north-south elongation of the error ellipses.  The origin of this
unexpected behavior can be seen in Figure~\ref{fig:lc}, which shows
that the satellite observations do not begin until a few days before
peak, which invalidates the underlying assumption of approximately
symmetric coverage about $t_0$ that was the basis of the statistical
independence of $t_0$ and $u_0$ for each observatory, and thus of
$\Delta t_0$ and $\Delta u_0$.  Instead, this strong asymmetry in
coverage leads to a strong correlation between $t_{0,\rm sat}$ and
$u_{0,\rm sat}$.  Indeed, such correlations can become even more
pronounced for the case that the satellite observations begin long
after $t_0$, in which case they can lead to large arcs on the
$\bpi_\e$ plane \citep{gould19}.  See \citet{kojima1c} for a dramatic
example of such an arc, which was nevertheless resolved by VLTI
interferometry.

Third, in this case the ambiguity between the two $\pi_{\e,\rm small}$
solutions is only marginally resolved because the directional
constraint (blue ray) for the ``wrong'' solution (right-middle panel)
passes close enough to the $\chi^2$ minimum that it must be
considered, while the difference between the minima is only
$\Delta\chi^2=7$.  Happily, and as anticipated, the magnitude $\pi_\e
= |\bpi_\e|$ is essentially the same for both solutions, so this
ambiguity does not strongly affect the mass and distance estimates,
but it does impact the directional information.  Hence, this event is
one that would profit from a second interferometric observation, which
would definitively resolve this ambiguity.  In
Section~\ref{sec:interferometry1}, I will discuss whether the decision
to undertake such a second observation could be made sufficiently
quickly. For now I note that even for this event, which is near the
margin of a reliable BH mass measurement (and was chosen as an example
for that reason), the chance of even marginal survival of the
alternate $\pi_{\e,\rm small}$ solution is only about 15\%.  To
understand this, the first point to note is that each event has equal
prior probability of having $u_0>0$ or $u_0<0$.  If I had simulated the
latter case (with all other parameters being the same), then the two
sets of contours in the middle row would have been approximately
flipped about the $x$-axis, but slightly shifted so that the $\chi^2$
minimum in the right panel exactly coincided with the actual parallax,
while the minimum of the left panel would be slightly displaced from
its flipped position.  The blue ray in the right panel would likewise
be flipped, so that it would pass exactly through the $\chi^2$
minimum.  However, the blue ray in the left panel would not pass
through its contours (now centered in the lower-left quadrant) because
it would pass through the (essentially empty) upper-right quadrant.
That is, the directional constraint for $u_0>0$ lies $90^\circ$
clockwise from the $u_0<0$ constraint (for interferometric
observations at $t=t_0+t_\eff$).  More generally, one should consider
an ensemble of random directions for $\Phi_\pi$ and both signs of
$u_0$.  At the small values of $\pi_\e$ probed by these diagrams, the
contours will retain their size, shape, and orientation as the ``true
solution'' is changed.  Hence, the probability of a marginally
resolved ambiguity is just the average angle subtended by the
$2\,\sigma$ contours as they are transported around the
$\pi_\e=0.0114$ circle (and averaged over both signs of $u_0$).  Note,
then, that the values of $\Phi_\pi$ with the smallest chance for a
surviving ambiguity (i.e., those for which the the long axis of the
contours point toward the origin), are also the ones with the greatest
errors in the $\pi_\e$ measurement.  The actual case shown in
Figure~\ref{fig:2comb1} is the opposite: the error ellipse is
transverse to directional constraint, which minimizes the $\pi_\e$
error, but maximizes the chance of survival of the ambiguity.

Finally, I note that a wider grid search rules out any additional (i.e.,
$\pi_{\e,\rm large}$) solutions.  I discuss the underlying reasons for this in
Section~\ref{sec:ground2}.

\subsubsection{Interferometric Decisions}
\label{sec:interferometry1}

\begin{figure}
\plotone{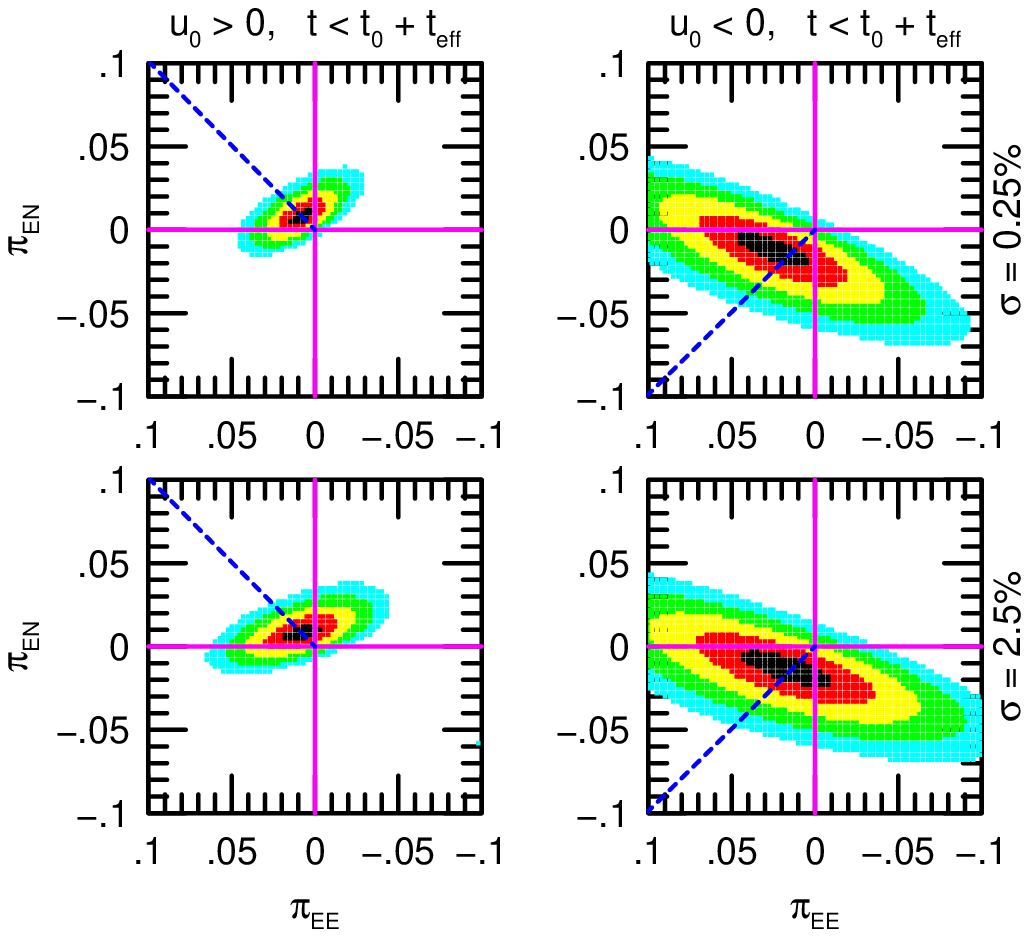}
\caption{$\bpi_\e$ contours from combined ground and satellite data that
  have been collected for Event~1 at the time of the interferometric decision,
  i.e., $t=t_0+t_\eff$.  The blue rays are shown as dashed because the
  interferometric result is not known when the decision is made to conduct
  this measurement, but is known when a second decision must be made on
  whether to obtain an additional interferometric measurement.
}
\label{fig:2comb1short}
\end{figure}

Figure~\ref{fig:2comb1short} shows the
$\bpi_\e$ contours at the time of the decisions about interferometric
observations, i.e., $t=t_0+t_\eff$, for both the weak ($\sigma=2.5\%$)
and ``realistic'' ($\sigma=0.25\%$) external flux constraints.  Note that
the scale is 5 times larger than the corresponding panels in
Figure~\ref{fig:2comb1}.  Also note that the blue rays that represent the
interferometric direction measurement are shown as dashed lines to
indicate the dual role of these diagrams:  Based on the contours alone
(no blue rays) one must decide whether the event is a sufficiently good
candidate to make the interferometric measurement.  Then, after the
measurement (so, with blue ray), one must decide whether it will be necessary
to make an additional measurement to break the directional degeneracy.

The first point is that including the ``realistic'' flux constraint does not
qualitatively improve the decision process.  In any case, I subsequently
refer to the ``realistic'' case, i.e., the upper panels.
Based on these, and at the $2\,\sigma$ level, $0.01\la\pi_\e\la 0.03$ and
$0.01\la\pi_\e\la 0.07$ for $u_0>0$ and $u_0<0$, respectively.
Thus, using Equation~(\ref{eqn:m-murel}) one finds that the $u_0<0$ solution is
consistent with an ordinary $M=1\,M_\odot$ star with proper motion
$\mu_\rel=3\,\masyr$.  Nevertheless, if true, the chance that the
much more tightly constrained $u_0>0$ contours would lie so close to the
origin is only about 25\%.  There might be additional information from
the FWHM$\sim 0.4^{\prime\prime}$ satellite images that ruled out main-sequence
lenses down to, say, $0.8\,M_\odot$, which could further inform the decision.
In the end, there would nevertheless be some risk that this was an ordinary
star, which could be mitigated by waiting to, e.g. $t=t_0+2\,t_\eff$, when
the source was fainter by 0.5 magnitudes.  The final decision would have to
take account of the feasibility of such future observations (based on
brightness, reference stars etc.), the time pressure on the instrument,
and the general level of risk that could be supported.

On the assumption that measurement was made, the result (intersection
of the blue rays with the contours), would imply that, at $2\,\sigma$,
$0.01\la\pi_\e\la 0.02$ and $0.01\la\pi_\e\la 0.04$ for the two cases.
Moreover, $\theta_\e=1.11\,\mas$ would now be measured, implying
$M=\theta_\e/\kappa\pi_\e\ga 3.4\,M_\odot$, i.e., a BH.  The fact that the
blue rays come close to both minima would indicate that it was unlikely
that the ambiguity would be decisively resolved by the full season of
observations and hence that an additional interferometric measurement
was warranted.  For example, by waiting just 3 days, the positional angle
of the major image relative to the minor image would rotate by $\sim 4^\circ$,
either clockwise for $u_0>0$ or counterclockwise for $u_0<0$.

Finally, I note that a comparison of the upper left panel of
Figure~\ref{fig:2comb1short} with the left panel of Figure~\ref{fig:ground1}
shows that the ground-only data plays almost no role in constraining the
combined-data contours.  For example, the black contours in
Figure~\ref{fig:ground1} intersect the $x$-axis over the range
$+0.05>\pi_{\e,E}>-0.04$, which is very large on the scale of the contours
in Figure~\ref{fig:2comb1short}.

\subsection{Event~2}
\label{sec:event2}

When analyzing Event~2, I mainly focus on understanding the differences
in results compared to Event~1.  These are quite dramatic given that
the only differences in event parameters are that $t_0$ is four
months later and the sign of $\pi_{\e,E}$ is reversed.

\subsubsection{Ground-Only Analysis}
\label{sec:ground2}

\begin{figure}
\plotone{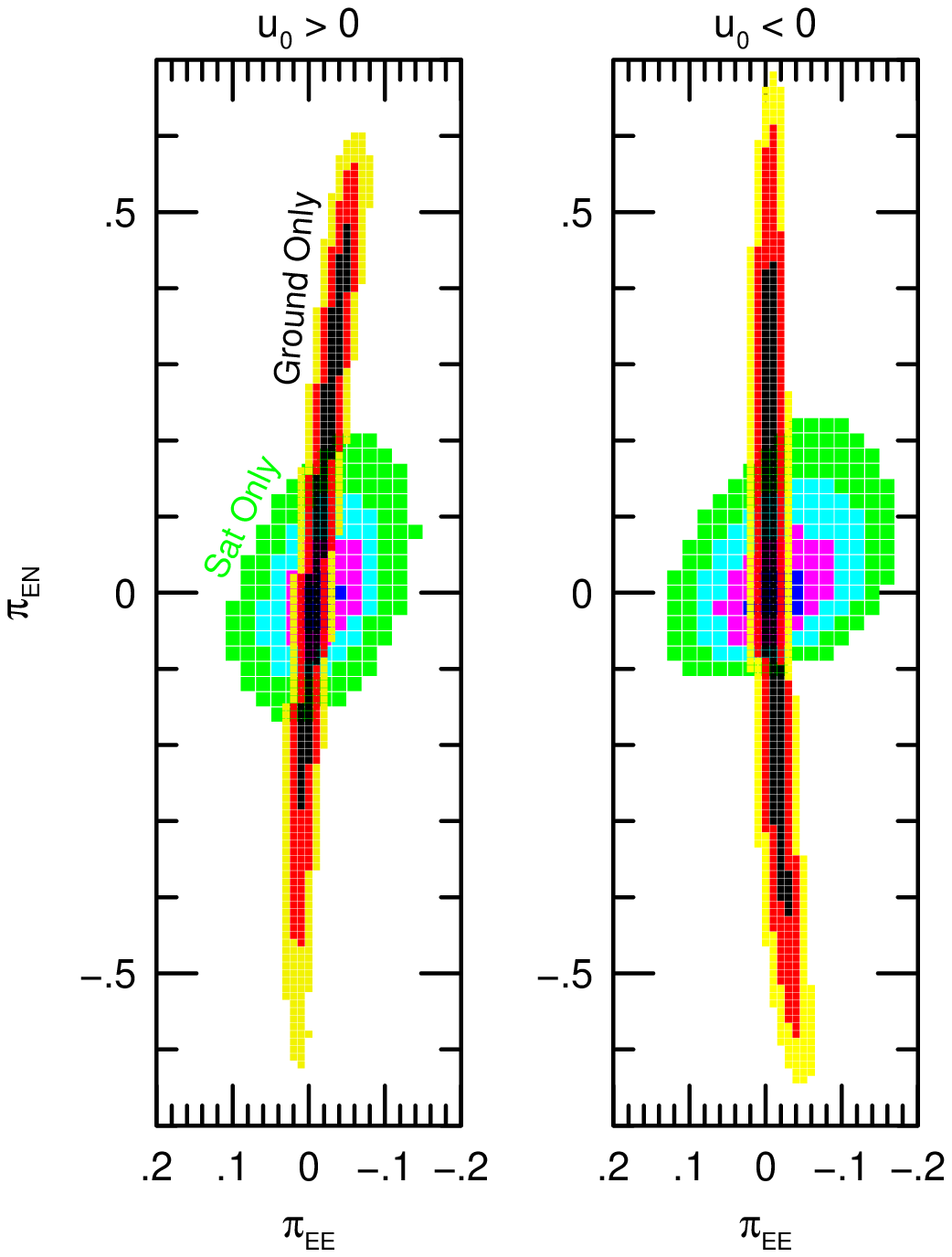}
\caption{$\bpi_\e$ contours for Event~2, based on
  ground-only (black, red, yellow) and satellite-only
  (blue, magenta, cyan, green) data.
}
\label{fig:overlap2}
\end{figure}

The nearly-linear (black, red, yellow) contours in Figure~\ref{fig:overlap2}
show the constraints on $\bpi_\e$ from ground-only data for Event~2.
These can be compared to the analogous contours for Event~1 that are
displayed in Figure~\ref{fig:ground1}.  The most striking difference is that
the constraints on $\pi_{\e,\parallel}$ (width of the contours) are weaker
by a factor $\sim 2$.  At least a partial explanation for this difference is
that $t_0$ for Event~2 is closer to the end of the ground microlensing
season than $t_0$ for Event~1 is to the beginning.  Therefore, the
asymmetry in the data with respect to $t_0$, from which the value of
$\pi_{\e,\parallel}$ is extracted, is less well traced by the data for Event~2.

A seemingly minor technical point is that the ground-only contours in
the $u_0>0$ panel of Figure~\ref{fig:overlap2} appear to be rotated
clockwise by about $16^\circ$.  Strictly speaking, as explained in
Section~\ref{sec:ground1}, they have been rotated counterclockwise by
$164^\circ$.  Applying the analysis given there,
I find that at $t_0$ for Event~2,
$(S_N,S_E)(t_0)=(-0.021,-0.952)\,\au$.  Hence, based on the
short-$t_\e$ formalism, we expect that the long axis of the contours
should be nearly straight and point $179^\circ$ (north through east).
As in Figure~\ref{fig:ground1}, both sets of ground-only contours are
curved, and as in that Figure, the tangents to the arcs at the 
best fit are not the same, being roughly $-4^\circ$ and $+2^\circ$ for
$u_0>0$ and $u_0,0$, respectively.  That is, their average is
$-1^\circ$, which (after a $180^\circ$ rotation) agrees with the
short-$t_\e$-framework prediction, as was also true for Event~1.

The reason that it is important to distinguish between a $16^\circ$ clockwise
rotation and a $164^\circ$ counterclockwise rotation is that this rotation helps
to understand how the $\pi_{\e,\rm large}$ solutions are automatically
excluded by the satellite data in most cases.

As discussed in Section~\ref{sec:simchars}, the satellite trails
Earth's orbit by about 82 days.  Thus, as seen from the satellite, the
Sun's position at $t_0$ is the same as it would be from Earth, but 82
days earlier, i.e., about 10 June, at which point
$(S_N,S_E)(t_0)=(-0.099,+0.140)\,\au$.  In fact, a precise calculation
based on the actual {\it Spitzer} orbit that is being used in these
calculations gives $(S_N,S_E)_{\rm sat}(t_{0,\rm ground})=(-0.082,+0.153)\,\au$.
However, for purposes of understanding the underlying issues, it is easier
to think about the more familiar Earth orbit near 10 June than the more
precise satellite orbit near 1 September.  At this time, the event is nearly
in opposition, which actually occurs on 18 June, when $S_E=0$.
During the $2|S_N|{\rm yr}/2\pi = 12\,$days from before to after opposition,
$\Psi_\parallel$ rotates by $90^\circ$.  Hence during these times, the
constant-acceleration formalism is not even remotely applicable for
Event~2, which has $t_\eff=21\,$days.  Therefore, we do not expect the
$\bpi_\e$ contours that are derived from satellite-only data to be long,
linear structures.  Nevertheless, we do expect that they will be
very different from the ground-only contours.  Figure~\ref{fig:overlap2},
which displays the satellite-only contours in (blue, magenta, cyan, green)
confirms this expectation.  In particular, there is no overlap between
the two sets of contours for $\pi_{\e,\rm large}\simeq 2 u_0(\au/D_\perp)\sim 0.6$
at any orientation $\Phi_\pi$.

While the particular contours shown in Figure~\ref{fig:overlap2} are
specific to Event~2, it will always be the case that if the satellite
trails (or leads) Earth by enough to make a reliable measurement of
$\Delta t_0$ and $\Delta u_0(+,+)$ for a BH lens, then the ground-only
and satellite-only contours will be substantially different,
particularly at large $\pi_\e$.  Hence, except for relatively
high-magnification events (for which $\pi_{\e,\rm large}$ is not very
big, and for which it may be the correct solution, even for BH
events), the combination of ground-only and satellite-only fits will,
by themselves almost always exclude the $\pi_{\e,\rm large}$
solutions.

Stepping back, once there are dense satellite data that well cover the peak
(which applies fully to Event~2, but only partially to Event~1),
there are actually three, quasi-independent, constraints on $\bpi_\e$:
first on $(\Delta t_0,\Delta u_0)$ from comparison of the \citet{pac86}
parameters of the ground-only and satellite-only fits; second from
the distortions induced by the accelerated motion of Earth on the
ground-based light curve; and third from the distortions induced on
the satellite-based light curve due to its acceleration.

For BH events of modest peak magnification, the first of these three
leaves a four-fold degeneracy, but provides nearly all the information about
each of the four local minimum, while the combination of the latter two
provides very little information about the two $\pi_{\e,\rm small}$ solutions,
but decisively excludes the two $\pi_{\e,\rm large}$ solutions.

\subsubsection{Ground+Satellite Analysis}
\label{sec:sat2}

\begin{figure}
\plotone{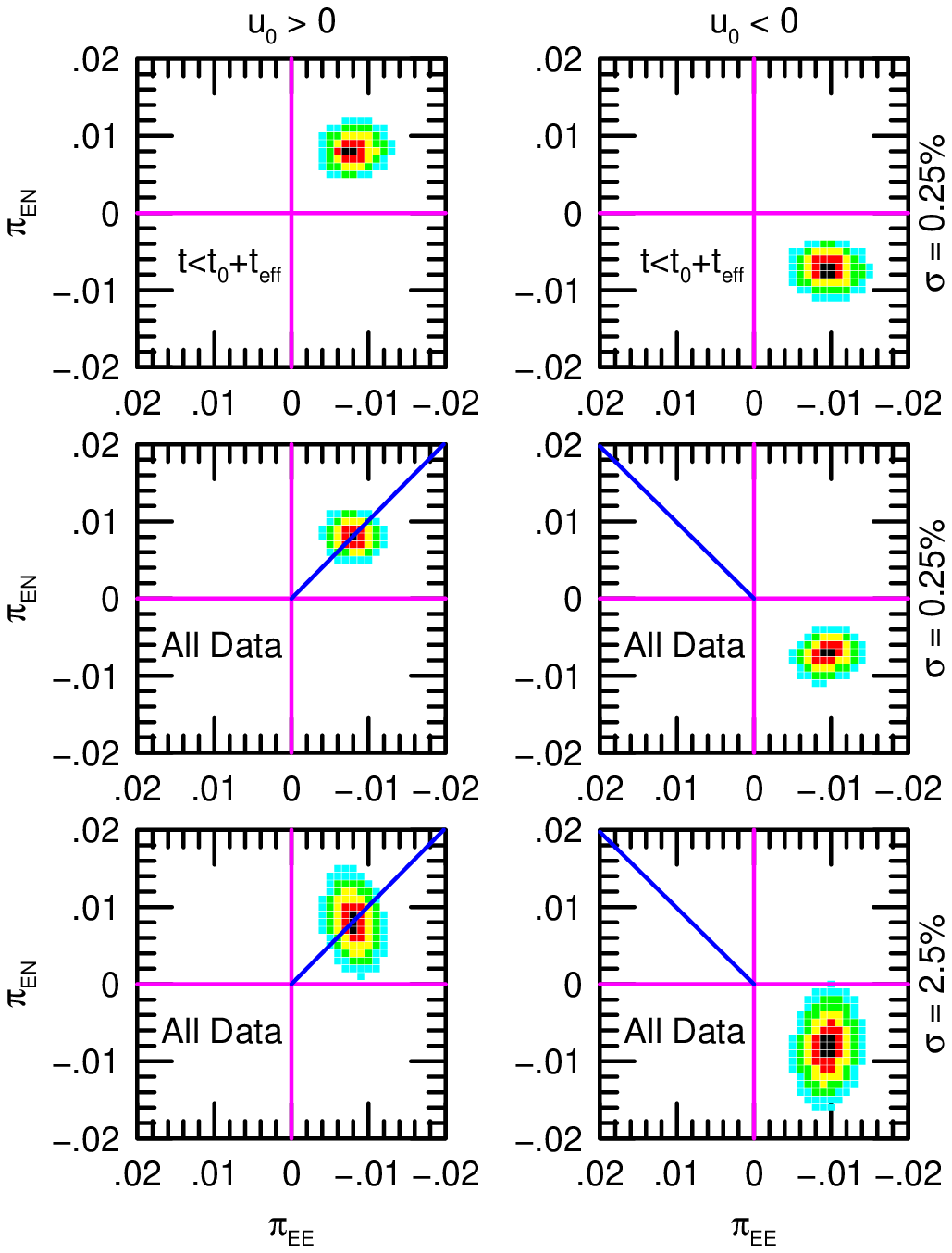}
\caption{The lower four panels show the $\bpi_\e$ contours from combining all
  of the ground and satellite data for Event~2. ``Realistic'' and very
  weak flux constraints are applied for the middle and lower panels,
  respectively.  These can be compared directly to the lower four panels
  of Figure~\ref{fig:2comb1}.  The two upper panels show the $\bpi_\e$ contours
  as of the time of the interferometric decision, i.e., $t=t_0 + t_\eff$.
}
\label{fig:2comb2}
\end{figure}

The lower four panels of Figure~\ref{fig:2comb2} show the $\bpi_\e$
contours for Event~2 for the cases of both the ``realistic'' and weak
flux constraints.  These can be directly compared to the lower four panels
of Figure~\ref{fig:2comb1}.  The differences are striking.  First,
the error ellipses are dramatically smaller in all four cases.  Second, the
interferometric measurement of the direction (blue rays) decisively
rules out the $u_0<0$ solution for Event~2.

The reasons for the smaller error ellipses can be understood directly
from Figure~\ref{fig:lc}.  From the top panel, we see that
$D_\perp(t_0)\sim 1\,\au$ for Event~2, whereas it is an order of
magnitude smaller for Event~1.  The result is that the Earth (blue)
and satellite (magenta) peaks are clearly separable by eye in both
time and amplitude for Event~2, while they appear
indistinguishable for Event~1.  (For Event~1, most of the information about
$\Delta t_0$ and $\Delta u_0$ comes from the falling part of the light
curve, over which $D_\perp$ is rapidly growing.  However, the
resulting differences in the two light curves do not yield features
that are discernible by eye.)\ \

The fact that the interferometric measurement rules out the alternate
$\pi_{\e,\rm small}$ solution is a matter of chance in two senses.
First, if the $u_0<0$ solution had been correct, then the interferometric
measurement would (of course) have resulted in a blue ray that passed
directly through the correct solution (which would be in the right panels).
However, in this case, the blue ray in the left panels would lie $90^\circ$
counterclockwise from this orientation, which would be approximately
its current location, i.e., passing directly through the (wrong)
$u_0>0$ solution.  Hence, this aspect is a 50-50 chance.

The second aspect of chance is much more one-sided.  The two blue rays
are displaced by $90^\circ$ because the interferometric observations occur
at $t_0 + t_\eff$.  At other times, the displacement would differ
by $2\phi$ where $\cot\phi=\delta t/t_\eff$, as described by
Equation~(\ref{eqn:phidef}).  On the other hand, the two solutions
are separated by $\sim 2\tan^{-1}|\pi_{\e,E}/\pi_{\e,N}|$ (where I have assumed
that ${\bf D}_\perp$ is approximately west).
In the examples that I have given,
with $|\pi_{\e,N}|\simeq |\pi_{\e,E}|$, this just happens to be about
$90^\circ$, but for other values it would be different.  For cases like
Event~2, with small error ellipses, it will be rare that these two
angles agree closely enough to leave an ambiguity in the $\pi_{\e,\rm small}$
solutions.

\subsubsection{Interferometric Decisions}
\label{sec:interferometry2}

The upper row of Figure~\ref{fig:2comb2} shows the $\bpi_\e$ contours
at the time of the interferometric decision, $t=t_0 + t_\eff$, assuming
the ``realistic'' flux constraint.  Both sets of contours are well-localized,
with $\pi_\e$ near its actual value of 0.0114.  Hence, the inferred
proper motion for a putative $1\,M_\odot$ lens would be
$\mu_\rel=\kappa M_\odot\pi_\e/t_\e \sim 0.5\,\masyr$, i.e., well beyond
any plausible threshold for a positive decision.
Comparing these contours to those 
in the middle row of Figure~\ref{fig:2comb1short} shows vastly higher
precision for Event~2 compared to Event~1.  As with the full-dataset comparison
that was discussed in Section~\ref{sec:sat2}, this is due partly to
larger $D_\perp$ near $t_0$ and partly to the longer data stream from
the satellite.

\section{{Discussion}
\label{sec:discuss}}

\subsection{{Extreme Applications, Challenges, and Orbits}
\label{sec:extreme}}

In Sections~\ref{sec:intro}--\ref{sec:bpie-measurement}, I have argued
for the feasibility of industrial-scale BH mass measurements and have
illustrated and further investigated this with two example events in
Section~\ref{sec:examples}.  By choosing a relatively large value of
the Earth-satellite separation, $D_{\rm sat}\sim 1.3\,\au$, I was able
to illustrate some of the most powerful applications as well as the most
severe challenges of the method.

Regarding the first, the $3\,\sigma$ (yellow) error contours in the middle-left
panel of Figure~\ref{fig:2comb2} are $\sim 5$ times smaller than the value
of $\pi_\e$, which means that a three-times smaller parallax could be measured
with $\sim 20\%$ precision.  For a typical bulge $\pi_\rel\sim 15\,\muas$, this
would correspond to an $M=128\,M_\odot$ BH.  Thus, if there were a bulge
population of very massive BHs, e.g., from an early era of massive star
formation, they could be reliably detected and accurately measured.
This sensitivity is made possible in part by the large
$D_{\rm sat}\sim 1.3\,\au$, which allowed $D_\perp\sim 1\,\au$ over the
peak of Event 2 (see Figure~\ref{fig:lc}).

However, the same wide separation also creates challenges.  Given the
$45^\circ$ Sun-exclusion angle, observations could be made only starting
on 23 April.  Engineering a smaller Sun-exclusion angle would be difficult,
and (unless it were at least $10^\circ$ smaller), there would be hardly
any benefit because $D_\perp$ would be very small at these times.
Measurements of $\bpi_\e$ could be made for events with $t_0$ before this date,
but they would have substantially larger $t_0$--$u_0$ correlations than
Event 1, and therefore substantially larger errors, which are already
dramatically larger than for Event 2.

If $D_{\rm sat}$ were reduced, say to $0.6\,\au$, both aspects, i.e., the power
and the challenges would be reduced.  For events peaking late in the
microlensing season, like Event~2, the errors would be increased by
a factor $\sim 2$ because $D_\perp$ would be roughly half as big as in
Event~2 as it was simulated in this paper.  On the other hand, the error bars
would be substantially improved for Event~1.  First, at $t_0$ (1 May),
the projected satellite separation would be $D_\perp\sim 0.5\,\au$.
i.e., about half of the value for Event~2 in Figure~\ref{fig:lc}.
Second, satellite observations could commence $\sim 35\,$days earlier,
i.e., about 19 March, yielding much better coverage of the rising
part of the satellite light curve.  Moreover, events could be observed
that peaked up to a month earlier than those that could be observed with
$D_{\rm sat}\sim 1.3\,\au$.

To fully assess all possible orbits would be well beyond the scope of
the present work.  I only raise this question to make clear that there
is an issue here.  One could also consider a variable orbit.  For
example, the satellite could start with a semi-major axis $a=1.03\,\au$,
so that it $d D_{\rm sat}/dt\sim 0.3\,\au$/yr and thus quickly achieve
$D_{\rm sat}\sim 0.6\,\au$ after 2 years.  Then the orbit could changed
to $a=1.01\,\au$, thus reducing the drift to
$d D_{\rm sat}/dt\sim 0.1\,\au$/yr.  This would make the satellite more
sensitive to BHs overall early in its mission, but more sensitive
to extreme BHs later on.

\subsection{{Comparison to \citet{gouldyee14}}
\label{sec:gy14}}

\citet{gouldyee14} proposed another approach to industrial-scale BH
mass measurements, which was further examined by \citet{kb222397}.
Here, I briefly describe this technique, which rests on a technology
that is still under development, and I compare its promise and
challenges to those of the approach presented here.

The \citet{gouldyee14} approach is to exploit the data stream arising
from a wide-field space-based microlensing planet survey, such as {\it
  Roman} (formerly {\it WFIRST}), {\it Euclid}, or {\it CSST}.  I will
specifically consider {\it Roman} because its microlensing-survey
characteristics are approximately known.  The basic idea is to combine
measurements of $\pi_{\e,\parallel}$ (derived from photometric
measurements of the source) and $\btheta_\e\equiv \bmu_\rel t_\e$
(derived from astrometric measurements of the source), to yield
$\pi_\e$ and $\theta_\e$, and so $M=\theta_\e/\kappa\pi_\e$ and
$\pi_\rel = \theta_\e \pi_\e$.  The {\it Roman} microlensing survey is
expected to monitor an area of $\Omega_{Roman}\sim 2.5\,{\rm deg}^2$ at a
cadence of $\Gamma\sim 4\,{\rm hr}^{-1}$ for 6 different 72-day
campaigns, each centered approximately on an equinox.  {\it Roman} has
a 2.4m mirror, and the project will be carried out using a broad $H$-band
filter.

Being centered on the equinoxes, the 72-day campaigns are well suited
to measuring 1-D parallaxes.  Moreover, it is plausible to expect
photometric errors that are near the photon limit and uncorrelated
in time.  The key issue, as already recognized by \citet{gouldyee14},
is whether the astrometric measurements will be similarly well-behaved.
If they are, then it is straightforward to calculate how well the magnitude
and direction of $\btheta_\e$ can be measured from the ``astrometric
microlensing'' technique that I described in Section~\ref{sec:intro}.

\citet{kb222397} carried out simulations of this approach for a
limited number of cases.  They characterized the conditions needed to
measure BH masses of bulge BHs with typical microlens parallaxes
$\pi_\e\sim 0.014$ as $u_0\la 0.4$, $\sigma_{\rm phot}<0.01$, and
$0<(t_0 - t_{\rm equinox}){\rm sgn}_{\rm equinox} < 36\,$day.  Here,
$t_{\rm equinox}$ is the time of equinox (about 21 March or 21 September) and
${\rm sgn}_{\rm equinox}$ is $+1$ for the vernal equinox and $-1$ for
the autumnal equinox.  They estimated that their photometric-precision limit
obtains for sources with $H_{\rm Vega}=20.4$ or $M_H=5.3$ (M0 dwarfs).
As \citet{kb222397}
themselves emphasized, their limited simulations are no substitute for a
systematic investigation of the \citet{gouldyee14} approach.  Nevertheless,
I will use them here as a rough guide because they constitute the only
study that has been carried out to date.

To compare to the approach here, I estimate (based on
Section~\ref{sec:examples}) a limit of $I=19.8$ (or $M_I=3.25$, assuming
typical $A_I=2.0$), $u_0<0.4$, and a range of $t_0$ covering 150 days
per season.  I estimate an ``effective area coverage'' of
$\Omega_\eff=25\,{\rm deg^2}$.  That is, even though the ground survey
covers $\sim 100\,{\rm deg^2}$, the event rate in the outlying fields
is substantially lower than in the {\it Roman} fields.  There are then
three factors that must be combined to compare the two approaches:
area, time interval, and luminosity function.  For the first,
$\Omega_\eff/\Omega_{Roman}=10$.  For the time interval, {\it Roman} has
$6\times 36=216\,$days of sensitivity, whereas the approach presented here
would have $5\times 150=750\,$days for a 5-yr mission, which yields a
ratio of $\sim 3.5$.  To evaluate the ratio of allowed regions of the
luminosity function, I first
convert the {\it Roman} limit to $M_I=7.0$, and then apply the
\citet{holtzman98} luminosity function to obtain a ratio of $1/12.5$.
Combining the three factors, I obtain a net advantage of a factor
$10\times 3.5/12.5\sim 2.8$ in favor of the method presented here.
I must emphasize
that my calculation is both crude and only applies to one type
of BH (namely bulge BHs).  So the main conclusion is that in terms of
numbers of BH mass measurements under ideal conditions, the two approaches
are comparable.

The main advantage of the \citet{gouldyee14} approach is that no
special effort is required to obtain the data: assuming a successful
launch and successful operations, the data stream will ``just appear''
in a publicly available archive.

The main disadvantage is that there is really no way to determine in
advance whether it will actually work, i.e., whether the astrometric
errors from many thousands of measurements can be treated as
independent and can therefore be combined according to naive
statistics (as assumed by \citealt{kb222397} in the above-cited
calculations).  At the $\sigma_{\rm phot}=0.01$ threshold, a single
measurement has an error of $\sigma_{\rm ast,1}=1\,\mas$.  We can
conceptually bin the data in one-week bins, which are satisfactory for
tracing the astrometric evolution.  Then, assuming uncorrelated
errors, each has an error of $\sigma_{\rm ast,week} = 40\,\muas$, or
about $4\times 10^{-4}$ pixels.  Hence, to achieve the overall
precision calculated by \citet{kb222397}, there cannot be unmodeled
systematic trends in the data (due to either instrumental or
astrophysical causes) at this level.  \citet{gouldyee14} give an overview
of the Rumsfeld triad of known-known, known-unknown, and unknown-unknown
systematic errors.  I refrain from rehearsing these here. I will just
note that the difficulty of interpreting the {\it Hubble Space Telescope (HST)}
astrometry for the case of OGLE-2011-BLG-0462
\citep{ob110462a,ob110462b,ob110462c} provides a sobering note of caution.
First, the resolution is 2 times better for {\it HST}
than {\it Roman}.  Second, the {\it HST} sampling of the PSF is somewhat
better.  Third, the astrometric precision required was almost an order
of magnitude less severe than what will be required for the \citet{gouldyee14}
approach.  Yet, \citet{ob110462a} and \citet{ob110462b} obtained
substantially different estimates of the direction of $\bpi_\e$
(or $\bmu_\rel$), which was only resolved (in favor of the first) by
additional work by \citet{ob110462c}.

Happily, as pointed out by \citet{gouldyee14}, there are robust methods
for determining whether or not the final results have been corrupted
by systematic errors, regardless of whether or not these systematic errors
can be identified.  However, these tests can only be carried out
after the fact.

My conclusion is that both approaches should be vigorously pursued.

\acknowledgments
I thank Subo Dong for valuable discussions.

\clearpage

\end{document}